\documentclass[traditabstract,longauth]{aa}

\usepackage{graphicx}
\usepackage{xcolor,soul}
\usepackage{svg}
\usepackage{longtable}
\usepackage{array}
\usepackage{lscape}
\usepackage{adjustbox}
\usepackage{xcolor}
\usepackage[toc,page]{appendix}

\usepackage{txfonts}
\usepackage{amsmath,siunitx}

\usepackage[breaklinks, colorlinks, citecolor=blue, linkcolor=blue]{hyperref}

\usepackage{xcolor}

\usepackage{orcidlink}
\usepackage{hyperref}

\providecommand{\bjdtdb}{\ensuremath{\rm {BJD_{TDB}}}}

\providecommand{\mst}{\ensuremath{\,{\rm M_\odot}}}
\providecommand{\rst}{\ensuremath{\,{\rm R_\odot}}}

\providecommand{\arcsec}{$^{\prime \prime}$}

\newcommand{\rebound}{\texttt{REBOUND}}
\newcommand{\whfast}{\texttt{WHFAST}}
\newcommand{\ias}{\texttt{IAS15}}

\graphicspath{{./}{figures/}}

\begin{document}

   \title{TOI-2458\,b: A mini-Neptune consistent with in situ hot Jupiter formation
   }

   \subtitle{}

   \author{J\'an \v{S}ubjak 
          \inst{1,2}\orcidlink{0000-0002-5313-9722}
    \and
          Davide Gandolfi
          \inst{3}\orcidlink{0000-0001-8627-9628}
    \and
          Elisa Goffo
          \inst{3,4}\orcidlink{0000-0001-9670-961X}
    \and
          David Rapetti
          \inst{5,6}\orcidlink{0000-0003-2196-6675}
    \and
          Dawid Jankowski
          \inst{7}\orcidlink{0009-0007-5863-9690}
    \and
          Toshiyuki Mizuki
          \inst{8,9}\orcidlink{0009-0002-9832-0004}
    \and
          Fei Dai
          \inst{10}\orcidlink{0000-0002-8958-0683}
    \and
          Luisa M. Serrano
          \inst{3}\orcidlink{0009-0006-4279-8032}
    \and
          Thomas G. Wilson
          \inst{11}\orcidlink{0000-0001-8749-1962}
    \and
          Krzysztof Go\'zdziewski
          \inst{7}\orcidlink{0000-0002-8705-1577}
    \and
          Grzegorz Nowak
          \inst{7}\orcidlink{0000-0002-7031-7754}
    \and
          Jon M. Jenkins
          \inst{5}\orcidlink{0000-0002-4715-9460}
    \and
          Joseph D. Twicken
          \inst{5,12}\orcidlink{0000-0002-6778-7552}
    \and
          Joshua N.\ Winn
          \inst{13}\orcidlink{0000-0002-4265-047X}
    \and
          Allyson Bieryla
          \inst{2}\orcidlink{0000-0001-6637-5401}
    \and
          David R. Ciardi
          \inst{14}\orcidlink{0000-0002-5741-3047}
    \and
          William D. Cochran
          \inst{15,16}\orcidlink{0000-0001-9662-3496}
    \and
          Karen A.\ Collins
          \inst{2}\orcidlink{0000-0001-6588-9574}
    \and
          Hans J. Deeg
          \inst{17,18}\orcidlink{0000-0003-0047-4241}
    \and
          Rafael A. Garc\'ia
          \inst{19}\orcidlink{0000-0002-8854-3776}
    \and
          Eike W. Guenther
          \inst{4}\orcidlink{0000-0002-9130-6747}
    \and
          Artie P. Hatzes
          \inst{4}\orcidlink{0000-0002-3404-8358}
    \and
          Petr Kab\'{a}th
          \inst{1}\orcidlink{0000-0002-1623-5352}
    \and
          Judith Korth
          \inst{20}\orcidlink{0000-0002-0076-6239}
    \and
          David W. Latham
          \inst{2}\orcidlink{0000-0001-9911-7388}
    \and
          John\,H.\,Livingston
          \inst{8,9,21}\orcidlink{0000-0002-4881-3620}
    \and
          Michael B. Lund
          \inst{14}\orcidlink{0000-0003-2527-1598}
    \and
          Savita Mathur
          \inst{17,18}\orcidlink{0000-0002-0129-0316}
    \and
          Norio Narita
          \inst{8,17,22}\orcidlink{0000-0001-8511-2981}
    \and
          Jaume Orell-Miquel
          \inst{17,18}\orcidlink{0000-0003-2066-8959}
    \and
          Enric Pall\'e
          \inst{17,18}\orcidlink{0000-0003-0987-1593}
    \and
          Carina M. Persson
          \inst{23}\orcidlink{0000-0003-1257-5146}
    \and
          Seth Redfield
          \inst{24}\orcidlink{0000-0003-3786-3486}
    \and
          Richard P. Schwarz
          \inst{2}\orcidlink{[0000-0001-8227-1020}
    \and
          David Watanabe
          \inst{25}\orcidlink{0000-0002-3555-8464}
    \and
          Carl Ziegler
          \inst{26}\orcidlink{0000-0002-0619-7639}
}

   \institute{Astronomical Institute, Czech Academy of Sciences, Fri{\v c}ova 298, 251 65, Ond\v{r}ejov, Czech Republic
         \and
         Center for Astrophysics ${\rm \mid}$ Harvard {\rm \&} Smithsonian, 60 Garden Street, Cambridge, MA 02138, USA
         \and
         Dipartimento di Fisica, Universit\'a degli Studi di Torino, via Pietro Giuria 1, I-10125, Torino, Italy
         \and
         Th\"uringer Landessternwarte Tautenburg, Sternwarte 5, 07778 Tautenburg, Germany
         \and
         NASA Ames Research Center, Moffett Field, CA 94035, USA
         \and
         Research Institute for Advanced Computer Science, Universities Space Research Association, Washington, DC 20024, USA
         \and
         Institute of Astronomy, Faculty of Physics, Astronomy and Informatics, Nicolaus Copernicus University, Grudzi\c{a}dzka 5, 87-100 Toru\'n, Poland
         \and
         Astrobiology Center of NINS, 2-21-1, Osawa, Mitaka, Tokyo, 181-8588, Japan
         \and
         National Astronomical Observatory of Japan, 2-21-2, Osawa, Mitaka, Tokyo 181-8588, Japan
         \and
         Institute for Astronomy, University of Hawai`i, 2680 Woodlawn Drive, Honolulu, HI 96822, USA
         \and
         Department of Physics, University of Warwick, Gibbet Hill Road, Coventry CV4 7AL, United Kingdom
         \and
         SETI Institute, Mountain View, CA  94043, USA
         \and
         Department of Astrophysical Sciences, Princeton University, Princeton, NJ 08544, USA
         \and
         NASA Exoplanet Science Institute-Caltech/IPAC, Pasadena, CA 91125, USA
         \and
         McDonald Observatory, The University of Texas, Austin Texas USA
         \and
         Center for Planetary Systems Habitability, The University of Texas, Austin Texas USA
         \and
         Instituto de Astrof\'\i sica de Canarias (IAC), C. V\'\i a L\'actea S/N, E-38205 La Laguna, Tenerife, Spain
         \and
         Universidad de La Laguna (ULL), Dept. de Astrof\'\i sica, E-38206 La Laguna, Tenerife, Spain
         \and
         Universit\'e Paris-Saclay, Universit\'e Paris Cit\'e, CEA, CNRS, AIM, 91191, Gif-sur-Yvette, France
         \and
         Lund Observatory, Division of Astrophysics, Department of Physics, Lund University, Box 118, 22100 Lund, Sweden
         \and
         Astronomical Science Program, Graduate University for Advanced Studies, SOKENDAI, 2-21-1, Osawa, Mitaka, Tokyo, 181-8588, Japan
         \and
         Komaba Institute for Science, The University of Tokyo, 3-8-1 Komaba, Meguro, Tokyo 153-8902, Japan
         \and
         Chalmers University of Technology, Department of Space, Earth and Environment, Onsala Space Observatory, SE-439 92 Onsala, Sweden
         \and
         Astronomy Department and Van Vleck Observatory, Wesleyan University, Middletown, CT 06459, USA
         \and
         Planetary Discoveries, Valencia CA 91354, USA
         \and
         Department of Physics, Engineering and Astronomy, Stephen F. Austin State University, 1936 North St, Nacogdoches, TX 75962, USA
        }

   \date{\today{}; \today{}}

 
\abstract{We report on the discovery and spectroscopic confirmation of TOI-2458\,b, a transiting mini-Neptune around an F-type star leaving the main-sequence with a mass of $M_\star=1.05 \pm 0.03$\,M$_{\odot}$, a radius of $R_\star=1.31 \pm 0.03$\,R$_{\odot}$, an effective temperature of $T_{\rm eff}=6005\pm50$\,K, and a metallicity of $-0.10\pm0.05$\,dex. By combining TESS photometry with high-resolution spectra acquired with the HARPS spectrograph, we found that the transiting planet has an orbital period of $\sim$3.74 days, a mass of $M_p=13.31\pm0.99$\,M$_{\oplus}$ and a radius of $R_p=2.83\pm0.20$\,R$_{\oplus}$. The host star TOI-2458 shows a short activity cycle of $\sim$54\,days revealed in the HARPS S-index and H$\alpha$ times series. We took the opportunity to investigate other F stars showing activity cycle periods comparable to that of TOI-2458 and found that they have shorter rotation periods than would be expected based on the gyrochronology predictions.
In addition, we determined TOI-2458's stellar inclination angle to be $i_\star\,=\,10.6_{-10.6}^{+13.3}$\,degrees.
We discuss that both phenomena (fast stellar rotation and planet orbit inclination) could be explained by in situ formation of a hot Jupiter interior to TOI-2458\,b. It is plausible that this hot Jupiter was recently engulfed by the star. Analysis of HARPS spectra has identified the presence of another planet with a period of $P\,=\,16.55\pm0.06$\,days and a minimum mass of $M_p \sin i=10.22\pm1.90$\,M$_{\oplus}$. Using dynamical stability analysis, we constrained the mass of this planet to the range $M_{c} \simeq  (10, 25)$\,M$_{\oplus}$.

}

\maketitle

%
%
%

\section{Introduction}\label{sec:introduction}

The first discovery of an exoplanet around a Sun-like star, 51~Peg\,b, was an unexpected breakthrough, given the planet's dissimilarity from any known planets within our Solar System \citep{Mayor95}. Subsequent population studies revealed that hot Jupiters exist around roughly 1\% of Sun-like stars \citep{Mayor11, Wright12}. Despite over two decades of research, the primary formation channel of hot Jupiters remains elusive. The formation mechanisms for hot Jupiters can be grouped into two categories, the first positing a slow orbital decay as a result of interactions with a protoplanetary disk \citep[e.g.,][]{Lin96, Ward97, Murray98}, and the second suggesting the tidal circularization of a planet on a highly eccentric orbit, created by gravitational interactions with other massive bodies \citep[e.g.,][]{Rasio96, Fabrycky07, Naoz11}. Over the past two decades, migration mechanisms have become an established theoretical paradigm.

The formation of hot Jupiters in situ represents a potential alternative to migration mechanisms \citep[e.g.,][]{Pollack96,Boss97,Batygin16,Hasegawa19,Poon21}. Several arguments have been proposed in support of this scenario. Firstly, the rate of disk-driven migration observed in simulations for giant planets appears to be independent of planetary mass \citep{Kley12}, suggesting that the physical properties of hot and cold giant planets should be similar on average. However, recent studies have shown that close-in and distant giant planet populations exhibit intrinsic differences, with hot giant planets generally having smaller masses \citep{Knutson14,Bryan16}. Secondly, in situ formation has previously been discounted due to a lack of sufficiently massive cores at small orbital radii \citep{Rafikov06}. However, radial velocity surveys and space-based transit photometry have shown that the inner regions of planetary systems are not typically empty. Instead, hot planets with a period of less than 100 days and a mass of less than 30\,M$_{\oplus}$ are highly prevalent, with $30\%-50\%$ of main-sequence stars in the solar neighborhood serving as hosts \citep[e.g.,][]{Dressing13,Petigura13,Winn15}.

Rapid gas accretion onto protoplanetary cores can occur in the innermost regions of the protoplanetary disk if the core's mass exceeds $M_{core} \geq 15$ M$_{\oplus}$ \citep{Batygin16}. However, it remains an open question whether there is sufficient gaseous material available within a few AU. Therefore, the initial population of close-in giant planets formed in situ is expected to consist of low-mass planets occupying a similar orbital range. Considering the high occurrence of low-mass planets, it is reasonable to assume that in situ-formed giant planets are often accompanied by low-mass planets on similar orbits. This assumption leads to dynamical evolution that has observable implications \citep{Batygin16}. Recent observations from space-based observatories and radial velocity follow-up measurements of known transiting hot Jupiters have led to the detection of new low-mass planetary companions to hot Jupiters. Examples of such systems include WASP-47 \citep{Becker15}, Kepler-730 \citep{Canas19}, TOI-1130 \citep{Huang2020,Korth23}, WASP-132 \citep{Hord22}, TOI-2000 \citep{Sha23}, and WASP-148 \citep{Hebrard20}. These new discoveries provide evidence that at least some hot Jupiters may have formed in situ. They also suggest that nearby low-mass planetary companions to stars hosting hot Jupiters may be more common than previously thought. Furthermore, such companions could be below the detection limits of many past surveys.

The majority of systems mentioned above exhibit low-mass companions on inner orbits to hot Jupiters, which is expected in the theory \citep{Batygin16}. In this case, the nodal recession of the inner orbit is always faster, meaning that a secular resonance can never be established. Hence, such planets are expected to remain in unperturbed, coplanar orbits. However, if low-mass companions exhibit on exterior orbits, the initially slower nodal regression of the orbit will accelerate as the inner body gains mass and the star spins down. Once affected by secular resonances, the orbit may become unstable or acquire high mutual inclination, making observations of both interior and exterior planets extremely challenging as, in this case, a non-transiting planet might orbit at a low inclination relative to the line of sight and produce only weak radial-velocity variations. Observation of a system with a transiting low-mass planet orbiting close to the poles of a star with non-transiting hot Jupiter would be an important piece of the puzzle addressing the in situ formation of hot Jupiters.


We have spectroscopically confirmed a transiting planet candidate from TESS TOI-2458\,b as a mini-Neptune planet orbiting an F-type star. Our paper includes a description of the observations in Section \ref{sec:observations}, data analysis in Section \ref{sec:analysis}, a discussion in Section \ref{sec:discussion}, and a summary of our results in Section \ref{sec:summary}.


%
%

\section{Observations}\label{sec:observations}

\subsection{TESS light curves}\label{sec:TESS}

The Transiting Exoplanet Survey Satellite \citep[TESS;][]{Ricker15} conducted a photometric monitoring campaign of TOI-2458 (TIC 449197831) in sector 32 from November 20, 2020, to December 16, 2020, employing a 120-second time sampling. The resulting data for TOI-2458 are publicly available on the Mikulski Archive for Space Telescopes (MAST)\footnote{\url{https://mast.stsci.edu/portal/Mashup/Clients/Mast/Portal.html}.}. These data were processed by the TESS Science Processing Operations Center \citep[SPOC;][]{Jenkins16} located at the NASA Ames Research Center. The SPOC conducted a transit search of the light curve on January 8, 2021, with an adaptive, noise-compensating matched filter \citep{Jenkins02,Jenkins10,Jenkins20}, producing a Threshold Crossing Event (TCE) with 3.74 d period for which an initial limb-darkened transit model was fitted \citep{Li19} and a suite of diagnostic tests were conducted to help assess the planetary nature of the signal \citep{Twicken18}. The TESS Science Office reviewed the vetting information and issued an alert for TOI 2458.01.01 on February 03, 2021 \citep{Guerrero21}.

The light curve (LC) derived from TESS sector 32 was obtained from the MAST archive using the {\tt lightkurve} software package \citep{Lightkurve18}. We used the PDCSAP LC \citep{Smith12, Stumpe2012, Stumpe2014} processed by the SPOC pipeline, which also removes systematic errors associated with the instrument. The Pixel Response Functions (PRFs) were employed to investigate the crowding, and a correction for crowding was incorporated into the PDCSAP flux time series. Failure to incorporate such a correction may lead to an underestimation of the planet's radius. In addition, the SPOC Data Validation difference image centroid offset analysis \citep{Twicken18} locates the source of the transit signature within $3.0\pm2.7$ arcsec of the target star. This excludes all TIC objects other than TOI-2458 as potential sources of the transit signature.

\begin{figure*}
\centering
\includegraphics[width=1.0\textwidth,height=0.7\textwidth]{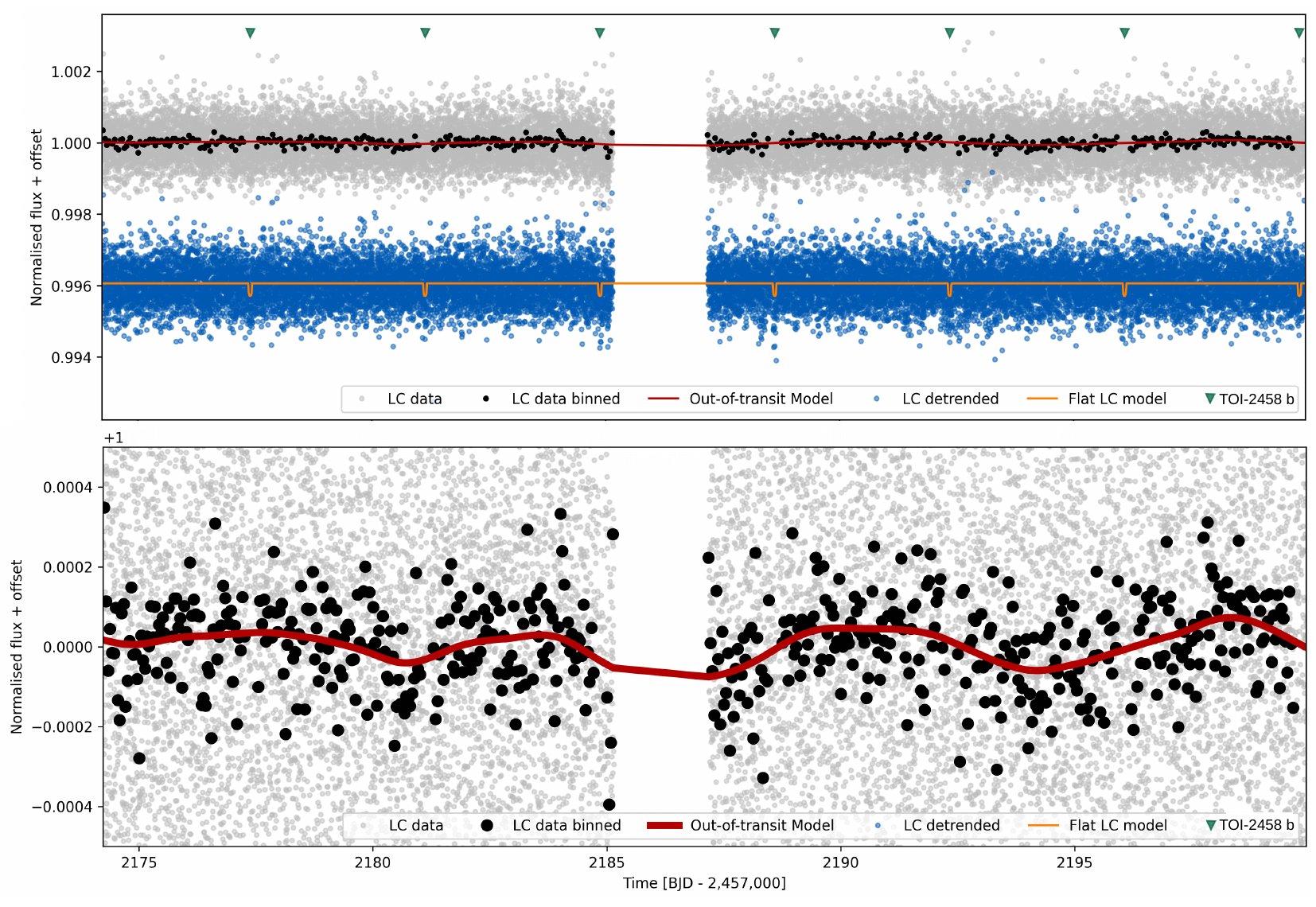}
\caption{SPOC PDCSAP LC from TESS sector 32 for TOI-2458 downloaded with the {\tt lightkurve} tool. Grey points represent the TESS observations, black points are binned TESS data, while red lines correspond to the out-of-transit GP models created with {\tt citlalicue} to capture the variability in the LC. Datasets were divided by these models, leading to a flattened TESS LC (blue points) with the transit model (orange line). Green triangles indicate the positions of transits. The lower panel is a zoom-in of the upper panel highlighting the out-of-transit GP model.}
\label{fig:tess_lc}
\end{figure*}

We utilized the Python package, {\tt citlalicue} \citep{Barragan22}, to remove residual stellar variability from the PDCSAP LCs. Combining a Gaussian process regression with transit models computed via the {\tt pytransit} code \citep{Parviainen15}, {\tt citlalicue} generated a model comprising both the light curve variability and the transits. We subtracted the variability part and produced a flattened LC containing only the transit photometric variations. We detected 7 transits. The LCs before and after the process is shown in Fig.~\ref{fig:tess_lc}.

\begin{figure}
\centering
\includegraphics[width=0.46\textwidth, trim= {0.0cm 0.0cm 0.0cm 0.0cm}]{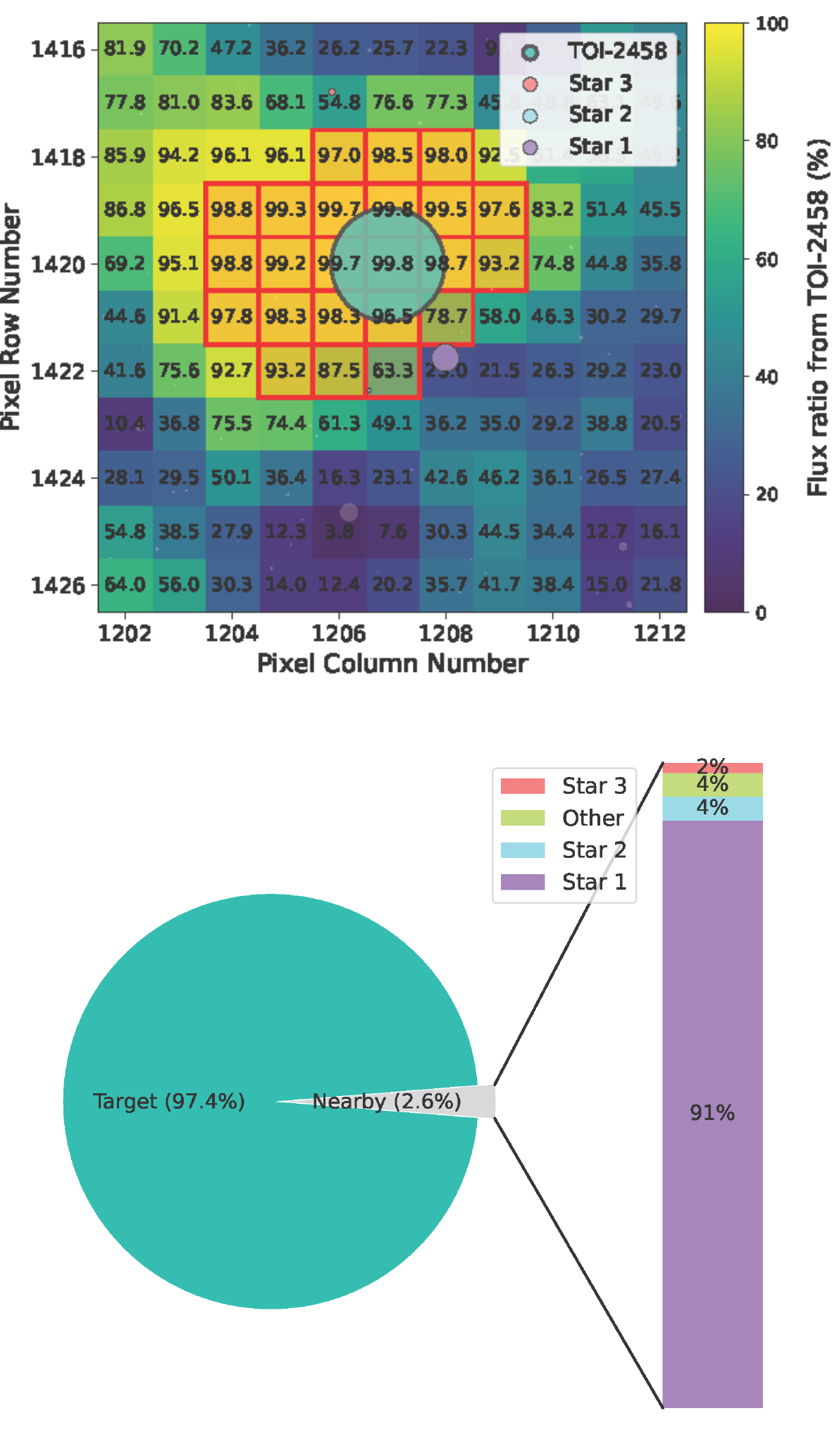}
\caption{Top: Heatmap illustrating the percentage of the flux from TOI-2458 that falls within each pixel of the TESS TPF image using Gaia DR2/DR3 catalogs. Bottom: Pie chart representing the flux from the target and nearby stars inside the photometric aperture. Star\,1 (Gaia ID 3233785163260952192) is the primary source of contamination.} \label{fig:field_image}
\end{figure}

To compare the Gaia DR2/DR3 catalogs with the TESS target pixel file (TPF), we utilized the {\tt TESS-cont} tool \citep{Castro-Gonzalez24}. This tool is used to detect any potential stars that could dilute the TESS photometry. Figure \ref{fig:field_image} presents a heatmap that illustrates the percentage of flux from the target star that falls within each pixel. As shown, within the aperture, TOI-2458 contributes between 63\% and 100\% of the total flux, depending on the specific pixel analyzed. We found that only 2.6\% of the flux in total originates from nearby sources, indicating negligible contamination, which has already been corrected in the PDCSAP. It is worth noting the presence of the dominant contaminating source Gaia ID 3233785163260952192, which is relatively close to the aperture used, although it is three magnitudes fainter.

\begin{table}
 \centering
 \caption[]{Basic parameters for TOI-2458.}
 \label{tab:basicpar}
 \begin{adjustbox}{width=0.48\textwidth}
    \begin{tabular}{lccr}
        \hline
	\hline
	Parameter & Description & Value & Source\\
        \hline
    $\alpha_{\rm J2000}$ & Right Ascension (RA) & 05:18:29.17 & 1\\
    $\delta_{\rm J2000}$ & Declination (Dec) & 01:15:13.92 & 1\\
    \smallskip\\
    $V_T$ & Tycho-2 $V_T$ mag & $9.268 \pm 0.020$ & 2\\
    $B_T$ & Tycho-2 $B_T$ mag & $9.900 \pm 0.028$ & 2\\
    $G$ & Gaia $G$ mag & $9.105 \pm 0.003$ & 1\\
    $B_P$ & Gaia $B_P$ mag & $9.384 \pm 0.003$ & 1\\
    $R_P$ & Gaia $R_P$ mag & $8.659 \pm 0.003$ & 1\\
    $T$ & TESS $T$ mag & $8.710 \pm 0.006$ & 3\\
    $J$ & 2MASS $J$ mag & $8.196 \pm 0.020$ & 4\\
    $H$ & 2MASS $H$ mag & $7.990 \pm 0.044$ & 4\\
    $K_S$ & 2MASS $K_S$ mag & $7.891 \pm 0.023$ & 4\\
    $WISE1$ & WISE1 mag & $7.827 \pm 0.028$ & 5\\
    $WISE2$ & WISE2 mag & $7.872 \pm 0.019$ & 5\\
    \smallskip\\
    $\mu_\alpha \cos{\delta}$ & PM in RA (mas/yr) & $-3.023 \pm 0.016$ & 1\\
    $\mu_{\rm \delta}$ & PM in DEC (mas/yr) & $-99.207 \pm 0.012$ & 1\\
    $\pi$ & Parallax (mas) & $9.027 \pm 0.017$ & 1\\
    \hline
    \multicolumn{3}{l}{Other identifiers:} \\
    \multicolumn{3}{l}{HD 34562} \\
    \multicolumn{3}{l}{TIC 449197831} \\
    \multicolumn{3}{l}{TYC 0100-00856-1} \\
    \multicolumn{3}{l}{2MASS J05182917+0115139} \\
    \multicolumn{3}{l}{Gaia DR2 3233785128901216128} \\
    \hline
    \hline
    \end{tabular}
    \end{adjustbox}
    \smallskip\\
    References: 1 - \citet{Gaia21}; \\ 2 - \citet{Hog00}; 3 - \citet{Stassun18}; \\ 4 - \citet{Skrutskie06}; 5 - \citet{Wright10}
\end{table}

\subsubsection{LCOGT Photometric Follow-up}

We observed a full transit window of TOI-2458 continuously for 260 minutes in Pan-STARRS $z$-short band on UTC 2021 February 14 from the Las Cumbres Observatory Global Telescope \citep[LCOGT;][]{Brown:2013} 1-m network node at Cerro Tololo Inter-American Observatory in Chile (CTIO). The $4096\times4096$ LCOGT SINISTRO cameras have an image scale of $0\farcs389$ per pixel, resulting in a $26\arcmin\times26\arcmin$ field of view. The images were calibrated by the standard LCOGT {\tt BANZAI} pipeline \citep{McCully:2018} and differential photometric data were extracted using {\tt AstroImageJ} \citep{Collins:2017}.

The TOI-2458 SPOC pipeline transit depth of 310\,ppm is generally too shallow to detect with ground-based observations reliably, so we instead checked for possible nearby eclipsing binaries (NEBs) that could be contaminating the TESS photometric aperture and causing the \textit{TESS} detection. To account for possible contamination from the wings of neighboring star PSFs, we searched for NEBs out to $2\farcm5$ from TOI-2458. If fully blended in the SPOC aperture, a neighboring star that is fainter than the target star by 8.8 magnitudes in TESS band could produce the SPOC-reported flux deficit at mid-transit (assuming a 100\% eclipse). To account for possible TESS magnitude uncertainties and possible delta-magnitude differences between TESS-band and Pan-STARRS $z$-short band, we included an extra 0.5 magnitudes fainter (down to \textit{TESS}-band magnitude 18.0). We calculated the RMS of each of the 33 nearby star light curves (binned in 10 minute bins) that meet our search criteria and find that the values are smaller by at least a factor of 5 compared to the required NEB depth in each respective star. We then visually inspected each neighboring star's light curve to ensure no obvious eclipse-like signal. Our analysis ruled out an NEB blend as the cause of the SPOC pipeline TOI-2458 detection in the \textit{TESS} data. All LCOGT light curve data are available on the {\tt EXOFOP-TESS} website.\footnote{\href{https://exofop.ipac.caltech.edu/tess/target.php?id=449197831}{https://exofop.ipac.caltech.edu/tess/target.php?id=449197831}}

\subsection{Contamination from nearby sources}\label{sec:ao_image}

To eliminate any potential sources of dilution within the Gaia separation limit of 0.4\arcsec, we obtained high-resolution images using adaptive optics and speckle imaging of TOI-2458. This was done as part of follow-up observations that were coordinated through the TESS Follow-up Observing Program (TFOP) High-Resolution Imaging Sub-Group 3 (SG3).

\begin{figure*}
\centering
\includegraphics[width=1.0\textwidth]{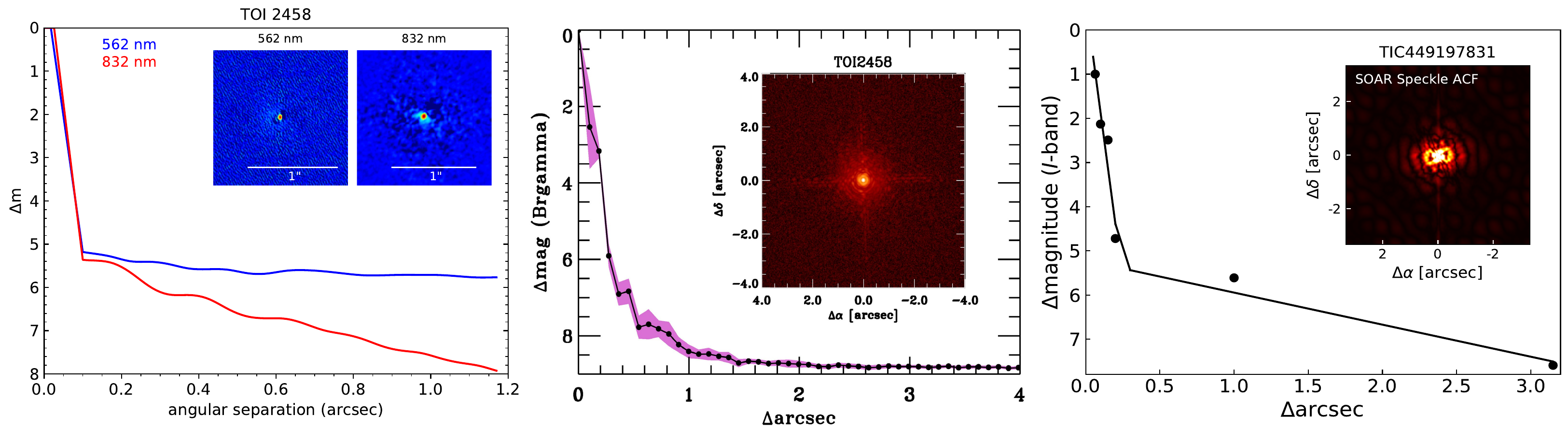}
\caption{From left to right: Alopeke contrast curve for 562\,nm and 832\,nm bands with a $1.2\arcsec \times 1.2\arcsec$ reconstructed image of the field. PHARO contrast curve for Brgamma band with a $8\arcsec \times 8\arcsec$ reconstructed image of the field. SOAR contrast curve for Cousins $I$ band with a $6\arcsec \times 6\arcsec$ reconstructed image of the field.} \label{fig:speckle_image}
\end{figure*}

On October 16, 2021, TOI-2458 was observed through the Alopeke speckle imager \citep{Scott18}, which is mounted on the 8.1\,m Gemini-North. The Alopeke system employs high-speed iXon Ultra 888 back-illuminated Electron Multiplying CCDs (EMCCDs) to simultaneously capture data in two bands centered around 562\,nm and 832\,nm. The data were processed using the procedures outlined in \citet{Howell11}. The final reconstructed image, displayed in Fig. \ref{fig:speckle_image}, attains a contrast of ${\Delta}mag$ = 6.60 at a separation of 0.5\arcsec in the 832\,nm band and ${\Delta}mag$ = 5.63 at a separation of 0.5\arcsec in the 562\,nm band. A contrast of ${\Delta}mag$\,$\sim5.50$ is achieved in both bands at a separation of 0.2\arcsec\, (22\,au). The estimated point spread function (PSF) is 0.02\arcsec wide.

On February 24, 2021, observation of TOI-2458 was conducted using the Palomar High Angular Resolution Observer \citep[PHARO;][]{Hayward01} with the JPL Palomar Adaptive Optics System, mounted on the 5.0\,m Hale telescope. PHARO uses a $1024 \times 1024$ HAWAII HgCdTe detector to capture images in the 1 to 2.5\,$\mu$m range. During the observation, a Br$\gamma$ filter was used. The reconstructed image, as shown in Fig. \ref{fig:speckle_image}, achieves a contrast of ${\Delta}mag$ = 6.83 at a separation of $0.5\arcsec$ with an estimated PSF of 0.09\arcsec width. We found no companion with a contrast of ${\Delta}mag$ $\sim 7$ at projected angular separations ranging from 0.5 to $4.0\arcsec$ (55 to 443 au).

On February 27, 2021, the High-Resolution Camera \citep[HRCam;][]{Tokovinin08} speckle interferometry instrument was employed on the Southern Astrophysical Research (SOAR) 4.1m telescope to observe TOI-2458. The observation was made in the Cousins $I$ filter with a resolution of 36\,mas. The observation was carried out according to the observation strategy and data reduction procedures outlined in \cite{Tokovinin18} or \cite{Ziegler21}. The final reconstructed image achieves a contrast of ${\Delta}mag$=5.5 at a separation of 0.5\arcsec, and the estimated PSF is 0.064\,arcsec wide. Fig. \ref{fig:speckle_image} plots a visual representation of the final reconstructed image. We found no companion with a contrast of ${\Delta}mag$ $\sim 5.5$ at projected angular separations ranging from 0.5 to $3.0\arcsec$ (55 to 332 au).



\subsection{HARPS spectra}

We acquired 87 spectra of TOI-2458 using the High Accuracy Radial velocity Planet Searcher \citep[HARPS;][]{Mayor03} spectrograph, which is mounted on the ESO 3.6-m telescope located at La Silla observatory. The observation period spans 590 days, from 16 February 2022 to 28 September 2023. The HARPS instrument has a spectral resolving power of R\,$\approx$\,115,000 and covers wavelengths ranging from $\sim$380\,nm to $\sim$690\,nm. The observations of TOI-2458 were carried out as part of our follow-up programs of TESS transiting planets (IDs: 1102.C-0923, 106.21TJ.001, 110.2438.001, 111.254R.002; PIs: Gandolfi, Dai, Wilson). We set the exposure time to $1200-1800$\,s depending on weather conditions, seeing, and nightly schedule constraints. We extracted the spectra using the HARPS data reduction software \citep[DRS;][]{Lovis2007} available at the telescope. The signal-to-noise ratio (S/N) per pixel at 550\,nm on the extracted spectra ranges from 57 to 124. We utilized the { \tt TERRA} algorithm \citep{Anglada12} to extract relative radial velocities (RVs) and spectral activity indicators. The {\tt TERRA} algorithm achieves this by using a least-squares matching technique of each observed spectrum to a high signal-to-noise ratio template, which was derived from the same observations. The radial velocities were adjusted for barycentric motion, secular perspective acceleration, instrumental drift, and nightly zero points.

%
%

\section{Analysis} \label{sec:analysis}

\subsection{Modeling Stellar Parameters with iSpec and SpecMatch} \label{st_par}

We achieved an S/N of 600 per pixel at 550\,nm by co-adding all the high-resolution HARPS spectra that were corrected for RV shifts. The combined spectrum was then used to determine the stellar parameters of TOI-2458 via the Spectroscopy Made Easy radiative transfer code \citep[{\tt SME};][]{Valenti96,Piskunov17}, which is incorporated into {\tt iSpec} \citep{Blanco14,Blanco19}. Complementary to this, we modeled the spectrum with MARCS models of atmospheres \citep{Gustafsson08} and utilized version 5 of the GES atomic line list \citep{Heiter15}. The iSpec software calculates synthetic spectra and compares them to the observed spectrum by minimizing the chi-squared value between them using a nonlinear least-squares (Levenberg-Marquardt) fitting algorithm \citep{Markwardt09}.

In our study, we employed the spectrum region spanning between $490-580$\,nm to determine the effective temperature $T_{\rm eff}$, metallicity $\rm [Fe/H]$, and the projected stellar equatorial velocity $v\sin{i}$. We utilized the Bayesian parameter estimation code {\tt PARAM 1.5} \citep{DaSilva06,Rodrigues14,Rodrigues17} to compute the surface gravity $\log{g}$ from PARSEC isochrones \citep{Bressan12} using derived spectral parameters as well as available photometry (refer to Table \ref{tab:basicpar}) and Gaia DR3 parallax as inputs. The iterative procedure was performed multiple times to achieve convergence to the final parameter values. The final parameters obtained are presented in Table \ref{table:stellar_par}. The obtained values agree with those calculated by the Gaia DR3 using low-resolution BP/RP spectra\footnote{\url{https://gea.esac.esa.int/archive/documentation/GDR3/Data_analysis/chap_cu8par/sec_cu8par_apsis/ssec_cu8par_apsis_gspphot.html}}.

\begin{figure}
\centering
\includegraphics[width=0.46\textwidth, trim= {0.0cm 0.0cm 0.0cm 0.0cm}]{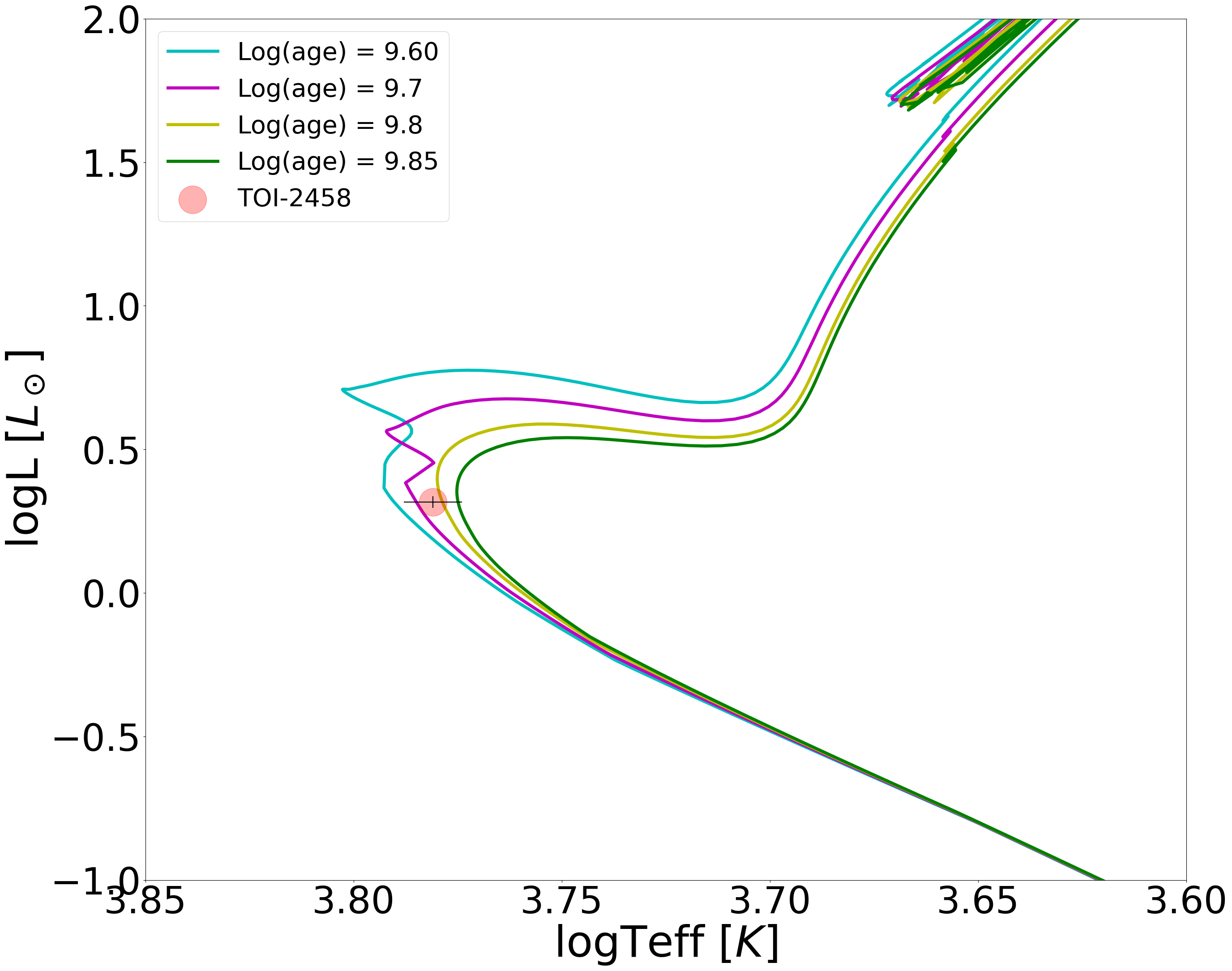}
\caption{Luminosity vs. effective temperature plot for the two-component model. Curves represent MIST isochrones for ages: 4.0\,Gyr (cyan), 5.0\,Gyr (purple), 6.3\,Gyr (yellow), 7.0\,Gyr (green), and for [Fe/H]\,=\,0.0. The red point represents the positions of TOI-2458 with its errorbars.} \label{fig:isochrones}
\end{figure}

The final parameters derived from several iterations were used as inputs to the {\tt PARAM 1.5} code once again to estimate the mass, radius, and age of the star. The star appears to be relatively old, with an estimated age of $5.7\pm0.9$\,Gyr. The age is also reported in Table \ref{table:stellar_par}. In Fig. \ref{fig:isochrones} we overplot the luminosity and effective temperature of TOI-2458 with the MIST stellar evolutionary tracks \citep{Choi16}. The spectral type of TOI-2458 was estimated via the latest version\footnote{\url{https://www.pas.rochester.edu/~emamajek/EEM_dwarf_UBVIJHK_colors_Teff.txt}.} of the empirical spectral type-color sequence proposed by \cite{Pecaut13}.

\begin{table}
 \centering
 \caption[]{Stellar parameters of TOI-2458.
 }
 \label{table:stellar_par}
\begin{adjustbox}{width=0.4\textwidth} 

	\begin{tabular}{lccccccr} 
		\hline
		\hline
& iSpec \& PARAM 1.5 analysis & SpecMatch \\
\hline
$\rm T_{eff}$ (K) & $6005 \pm 50$ & $6002 \pm 110$ \\
$[{\rm Fe/H}]$ (dex) & $-0.10 \pm 0.05$ & $0.05 \pm 0.09$ \\
$\log{g}$ (cgs) & $4.20 \pm 0.05$ & $4.28 \pm 0.12$ \\
$v_{\rm rot} \sin{i_\star} $ (km/s) & $2.5 \pm 2.5$ \\
$EW_{Li}$ (\AA) & 0.055 & \\
$M_\star$ ($\rm \mst$) & $1.05 \pm 0.03$ \\
$R_\star$ ($\rm \rst$) & $1.31 \pm 0.03$ \\
\hline
& ARIADNE analysis & \\
\hline
$\rm T_{eff} (K)$ & $6047 \pm 49$ \\
$[{\rm Fe/H}]$ & $-0.11 \pm 0.16$ \\
$\log{g}$ & $4.22 \pm 0.17$ \\
$L_\star$ (L$_{\odot}$) & $2.09 \pm 0.10$ \\
$R_\star$ ($\rm \rst$) & $1.32 \pm 0.03$ \\
\hline
& Other parameters & \\
\hline
$P_{rot}$ (days) & $8.9_{-2.5}^{+3.9} $ \\
Spectral type & F9--F9.5 \\
Age (Gyr) & $5.7_{-0.8}^{+0.9}$ \\
\hline
\hline
	\end{tabular}
\end{adjustbox}
\smallskip\\
\end{table}

\subsection{Stellar parameters with SpecMatch}\label{sec:SpecMatch}

As an additional validation of the derived stellar parameters, we used the HARPS co-added spectrum as an input to the {\tt SpecMatch} package \citep{Yee17}. The determination of stellar parameters by {\tt SpecMatch} involves a comparison of the observed spectrum with a library of well-characterized high signal-to-noise (S/N\,>\,400) HIRES spectra. All the derived parameters are documented in Table \ref{table:stellar_par} and are in agreement with the ones derived from {\tt iSpec}.

\subsection{SED analysis with ARIADNE}\label{sec:ARIADNE}

\begin{figure}
\centering
\includegraphics[width=0.46\textwidth, trim= {0.0cm 0.0cm 0.0cm 0.0cm}]{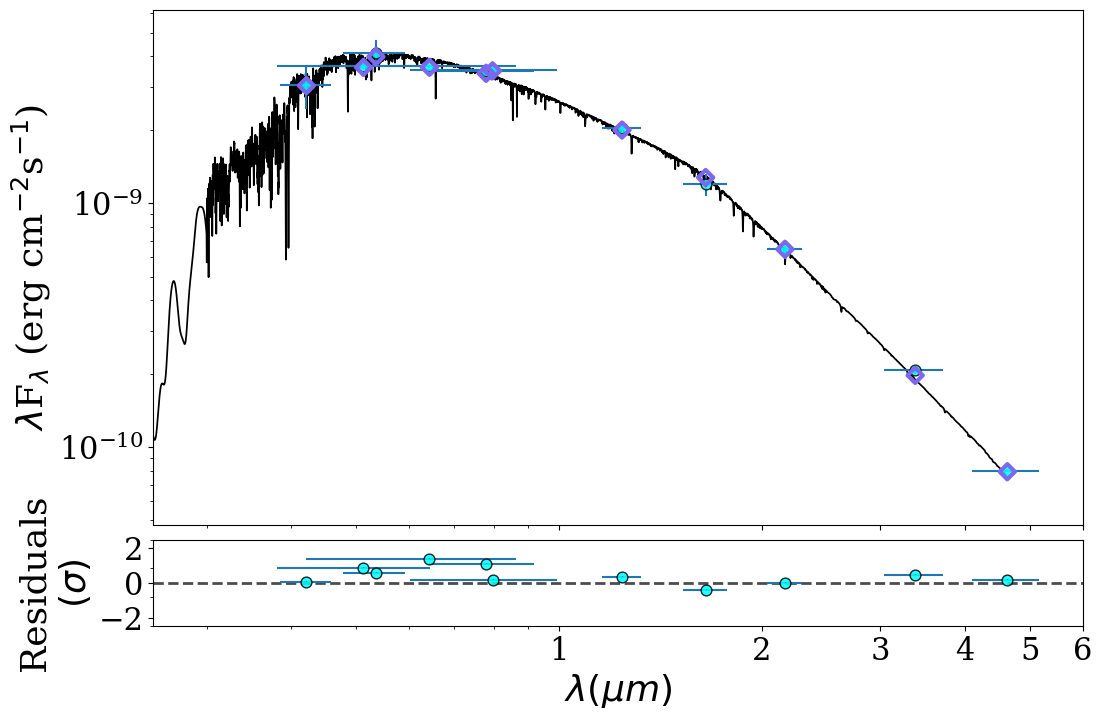}
\caption{SED fit for the TOI-2458 system. The blue points with vertical error bars show the observed catalog fluxes and uncertainties from Gaia, TESS, Tycho-2, 2MASS, and WISE, while the horizontal errorbars illustrate the width of the photometric band. For illustrative purposes, we overplot the best-matching Phoenix atmospheric model \citep{Husser2013}.} \label{fig:SED}
\end{figure}

We conducted an independent verification of the derived stellar parameters by modelling the Spectral Energy Distribution (SED) using the spectrAl eneRgy dIstribution bAyesian moDel averagiNg fittEr \citep[{\tt ARIADNE};][]{Vines22}. To determine the effective temperature $T_{\rm eff}$, surface gravity $\log{g}$, metallicity $\rm [Fe/H]$, and stellar luminosity and radius, we employed a combination of 3D dust Bayestar maps for interstellar extinction \citep{Green19} and four distinct models: Phoenix \citep{Husser2013}, BT-Settl \citep{Allard2012}, Kurucz \citep{Kurucz93}, and Ck04 \citep{Castelli2004}. Our priors for these parameters were based on results obtained from {\tt iSpec}. The derived final parameters, which are listed in Table \ref{table:stellar_par}, are in good agreement with previously derived parameters. The SED of TOI-2458 is shown in Fig.~\ref{fig:SED}, along with the best fitting Phoenix atmospheric model \citep{Husser2013}.

\subsection{Stellar rotation} \label{rotation}

The SPOC PDC pipeline module often eliminates a portion of the stellar variability. Therefore, we used the {\tt lightkurve} package to retrieve the TESS target pixel files from the MAST archive. Subsequently, a suitable aperture mask was selected to obtain a light curve, which was then normalized and corrected for outliers. To eliminate systematics from the data, we utilized the Pixel Level Decorrelation (PLD) method \citep{Deming15}. Data points within transits were excluded from the analysis. The SPOC SAP light curve peak-to-peak range is 1.9\% while the peak-to-peak amplitude of the oscillation recovered by PLD appears to be $\sim$$5.6x10^{-4}$, which is 2.9\% of the full range of the SAP light curve. To determine the stellar rotation, we employed the Gaussian Process (GP) Regression library, {\tt Celerite2}. In \citet{Foreman17,celerite2}, a detailed explanation of the library is provided. To account for variations in LCs that arise from heterogeneous surface features such as spots and plages, we used a rotational kernel function in the form of a sum of two stochastically-driven damped simple harmonic oscillators (SHOs) with periods of $P_{rot}$ and $P_{rot}/2$. This GP was successfully used in several papers studying stellar rotation \citep[e.g.,][]{David2019,Gillen20}. The power spectral density of both terms is:

\begin{figure*}
\centering
\includegraphics[width=1.0\textwidth]{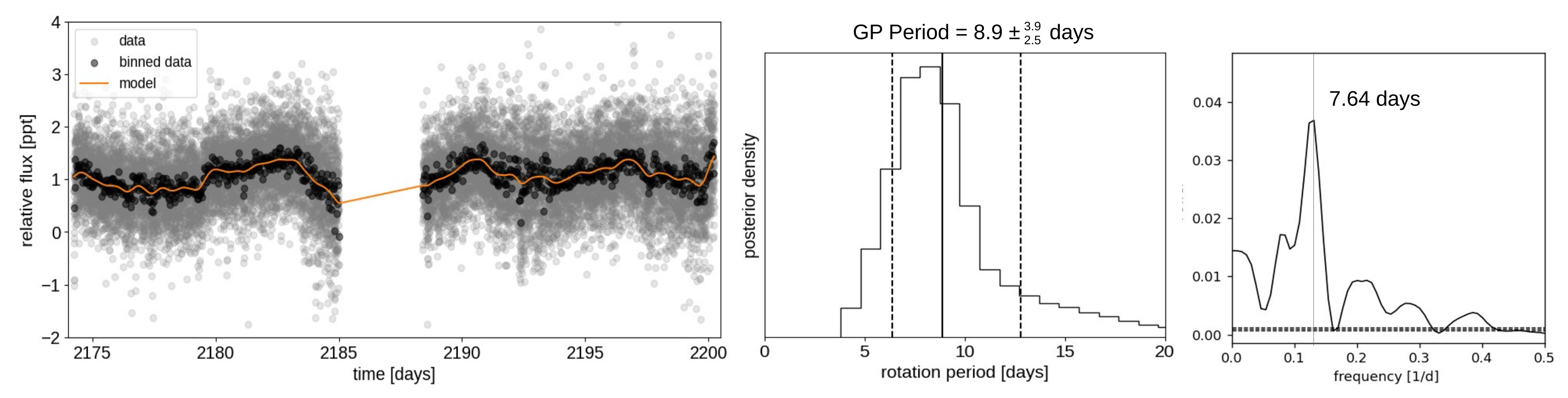}
\caption{Left: TESS PLD LC (gray points) with the MAP model prediction. It is important to note that the long trends and transits have been removed prior to analysis. The orange line shows the predictive mean. Middle: probability density of the rotation period. The 1$\sigma$ error bar is indicated with the dashed black lines. Right: GLS periodogram of the TESS PLD LC which was processed in accordance with the methods outlined in the text.} \label{fig:gp_plots}
\end{figure*}

\begin{equation}\label{eq_rot}
      S(\omega) = \sqrt{\frac{2}{\pi}}
         {\frac{S_{k}\omega^4_{k}}{(\omega^2-\omega^2_{k})^2+2\omega^2_{k}\omega^2/Q^2_k}},
   \end{equation}

\noindent
where, 
\begin{align}
      Q_1 = 1/2 + Q_0 + {\delta}Q, \\
      \omega_{1} = \frac{4{\pi}Q_1}{P_{rot}\sqrt{4Q^2_1-1}}, \\
      S_1 = \frac{\sigma^2}{(1+f)\omega_1Q_1}, \\
      Q_2 = 1/2 + Q_0, \\
      \omega_{2} = \frac{8{\pi}Q_2}{P_{rot}\sqrt{4Q^2_2-1}}, \\
      S_2 = \frac{f\sigma^2}{(1+f)\omega_2Q_2},
   \end{align}

where $\omega$ is the angular frequency, $S_k$ is the amplitude of the oscillation, $\omega_k$ is the undamped frequency, $Q_k$ is the quality factor, and $P_{rot}$ is the rotation period. The maximum a posteriori (MAP) parameters are estimated using the L-BFGS-B nonlinear optimization routine \citep{Byrd95,Zhu97}. To obtain the marginalized posterior distributions of free parameters, we ran a Markov Chain Monte Carlo (MCMC) analysis using {\tt emcee} \citep{Goodman10,Foreman13}. Our analysis yielded a derived rotational period of $P_{rot}=8.9_{-2.5}^{+3.9}$\,days, which is shown in Fig.~\ref{fig:gp_plots} as the probability density of $P_{rot}$ alongside the MAP model prediction.

We have also applied the generalized Lomb-Scargle (GLS) periodograms \citep{Zechmeister09} to the light curve obtained from TESS to verify the derived rotational period independently. Our analysis indicates a significant peak around 7.64 days, as illustrated in Fig. \ref{fig:gp_plots}. The stellar rotation signal observed in the GLS periodogram is consistent within error bars with the results obtained from the GP analysis.

\begin{figure}
\centering
\includegraphics[width=0.46\textwidth, trim= {0.0cm 0.0cm 0.0cm 0.0cm}]{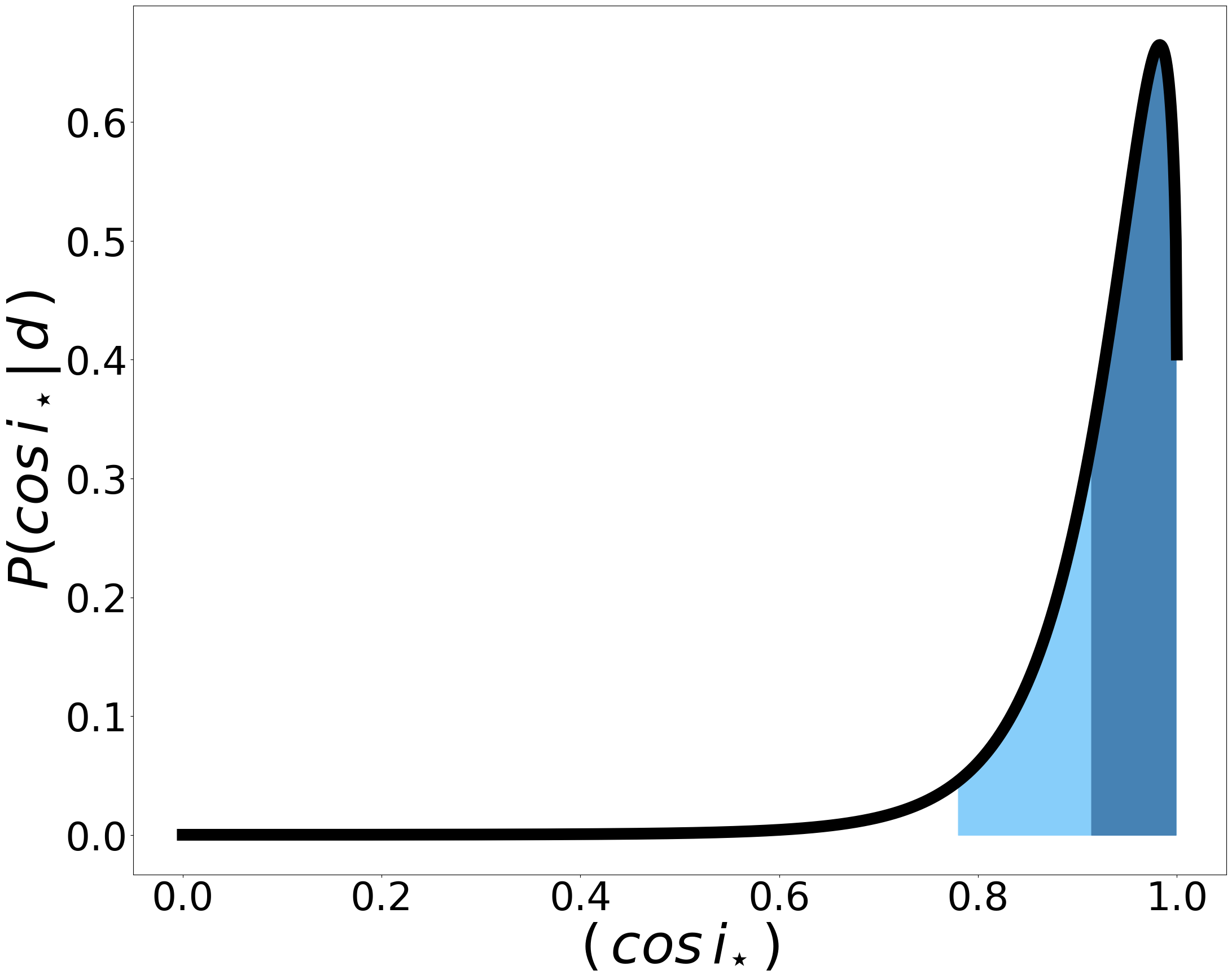}
\caption{Posterior distributions for line-of-sight stellar inclination in the form of cos\,i$_{\star}$. Blue shadow regions show 1-$\sigma$ and 2-$\sigma$ credible intervals.} \label{fig:inclination}
\end{figure}

\subsection{TESS variability validation} \label{variability}

The rotation period of TOI-2458 holds significant importance in further discussions regarding the system. To validate the variability in TESS LCs, we extracted cuts of TESS full-frame images around TOI-2458 and analyzed the variations in its neighboring stars. We used the same detrending vectors for each star to remove systematics. While none of the stars exhibit similar variability to TOI-2458, we still hesitate to rule out that we observe residual variations caused by systematic effects, given the suspicious phase that appears to be roughly synced with the spacecraft orbit. The WASP photometry has been able to exclude any variability of TOI-2458 down to a limit of 0.6\,mmag, which is insufficient to negate the variability observed in the TESS LC. Hence, the null detection in the WASP photometry is likely due to the small photometric amplitude of the signal. 

We investigated the robustness of the isolation of the stellar rotation signal by performing a series of injection and recovery experiments. We used the TESS Pixel Response Functions (PRFs) to inject sinusoidal signals modeled after the isolated signal, with a period of 7.5 days and an amplitude of 50 electrons/s, into the target pixel file (TPF) for TOI-2485 at six phases ($\ang{0}$, $\ang{45}$, $\ang{90}$, $\ang{135}$, $\ang{180}$, and $\ang{270}$) with respect to the stellar rotation signal. We also tested an injection with half of the amplitude at $\ang{90}$ with similar results. A customized PLD algorithm with an automatic corrector parameter optimization \citep{2024tsc3.confE..17R,rapetti} was then applied to both the original and the modified TPFs. The PLD-corrected light curves from the original TPFs were subtracted from those obtained from the injected TPFs to produce residual light curves that should only contain the injected stellar rotation signal. In general, PLD was able to recover the injected sinusoid at approximately the correct amplitude and phase, with some dependence on the phase angle with respect to long-term trends in the residual light curve and distortions in the recovered waveform near the edges of each orbit that were generally stronger in the first several days of the observations in each orbit. An alternative bespoke corrector based on that of \citet{Hedges21} was also investigated in this manner and provided comparable results. A third regression corrector utilizing the SPOC Cotrending Basis Vectors (CBVs) tended to overfit the data and not recover the injected signals, except for the case with a phase shift of $\ang{270}$. These results increase confidence in the interpretation of the 7.5-day signal as being intrinsic to the star. However, the ratio of the peak-to-peak range of the suspected stellar oscillation ($\sim$56\,ppm) compared to the peak-to-peak range of the systematic effects in the simple aperture light curve ($\sim$19,300\,ppm), along with the similarity in the oscillation period with those at which instrumental effects can occur, highlight the need for some caution with respect to this interpretation.

Additional high-precision photometry is required to confirm the rotation period of the star. Nevertheless, we have found a similar signal in the HARPS H$\alpha$ activity indicator, as discussed in Section \ref{sec:Freq}. Furthermore, in Section \ref{sec:activity}, we discuss that TOI-2458's rapid stellar rotation would not be surprising. It is imperative to confirm the star's rotation period for a more comprehensive validation of the system architecture. However, corresponding to our current best understanding, we elaborate on this potential scenario and assume that this is the star's rotation period for the remainder of this work.

Stellar inclination angle can be determined by combining of $P_{rot}$, $v \sin{i}$, and $R_\star$, yielding a value for TOI-2458 of $i_\star\,=\,10.6_{-10.6}^{+13.3}$\,degrees. The methodology employed in this analysis follows the procedure outlined in \cite{Masuda20} and implemented in \cite{Bowler23}. This approach correctly accounts for the correlation between stellar equatorial velocity and projected rotational velocity. The posterior distribution of stellar inclination can be viewed in Fig.\ref{fig:inclination}. 

\subsection{Age analysis}\label{sec:age}

In this section, we have employed various techniques to constrain the age of TOI-2458. Our analysis indicates that stellar isochrones (see Section \ref{st_par}) and all age indicators, except gyrochronology, are consistent with a relatively old age of $5.7\pm0.9$\,Gyr.

\subsubsection{Lithium EW}

The 6708\,\AA\ lithium line was employed to derive the equivalent width (EW) using the {\tt iSpec} tool. The EW of 0.055\,$\AA$ was determined by fitting the line with a Gaussian profile and calculating the area under the resulting curve. Subsequently, we compared the EW of lithium to that of well-established members of studied clusters from the literature. These are the Tuc-Hor young moving group \citep[$\sim$45 Myr;][]{Mentuch08}, the Pleiades \citep[$\sim$120 Myr;][]{Soderblom93}, M34 \citep[$\sim$220 Myr;][]{Jones97}, Ursa Major Group \citep[$\sim$400 Myr;][]{Soderblom1993}, Praesepe \citep[$\sim$650 Myr;][]{Soderblom993}, Hyades \citep[$\sim$650 Myr;][]{Soderblom90}, and M67 clusters \citep[$\sim$4 Gyr;][]{Jones99}. If necessary, we dereddened clusters using E(B-V) values obtained from \citet{Gaia2018}. Our findings reveal that TOI-2458 exhibits quite a high abundance of lithium. However, our measurements are still consistent with those obtained for the 4\,Gyr-old cluster M67.

\begin{figure}
\centering
\includegraphics[width=0.46\textwidth, trim= {0.0cm 0.0cm 0.0cm 0.0cm}]{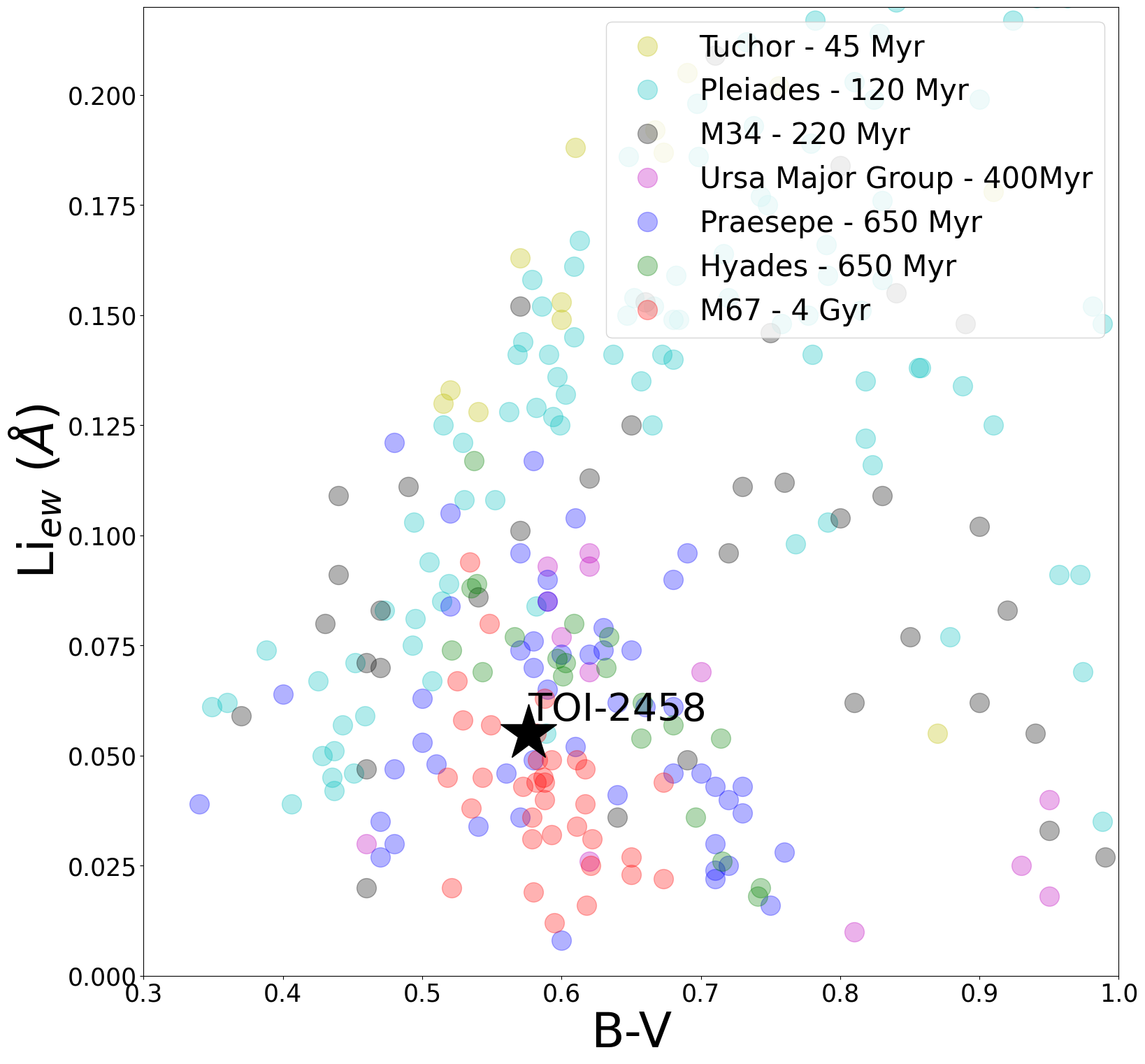}
\caption{Color vs. EW of lithium line Li\,6708 \AA{} of TOI-2458 together with members of well-studied clusters.} \label{fig:litium}
\end{figure}

\subsubsection{$R'_{HK}$ index}

To obtain a time-averaged measurement of the activity index of TOI-2458, $\log R'_{HK} = -4.98 \pm 0.06$, we utilized standard relations to convert the time-averaged $S$-index measurements from our spectroscopy \citep{Noyes84,Mittag13} using {\tt PyAstronomy\footnote{\url{https://pyastronomy.readthedocs.io/en/latest/pyaslDoc/aslDoc/sindex.html}.}}. The correlation between $R'_{HK}$ and $P_{rot}$ observed in numerous stars \citep[e.g.,][]{Lovis11,Mascareno15}. Utilizing the relationship established by \citet{Mascareno15}, it is possible to convert the value of $\log R'_{HK} = -4.98 \pm 0.06$ to a rotation period ranging from 23 to 40 days. This conversion indicates that the $\log R'_{HK}$ index for TOI-2458 is consistent with relatively advanced stellar age, and the expected rotation period is longer than observed. Based on the research conducted by \citet{Mamajek:2008}, the derived value of the $\log R'_{HK}$ index is consistent with the typical values observed in field stars.

\subsubsection{Gyrochronology}

\begin{figure}
\centering
\includegraphics[width=0.46\textwidth, trim= {0.0cm 0.0cm 0.0cm 0.0cm}]{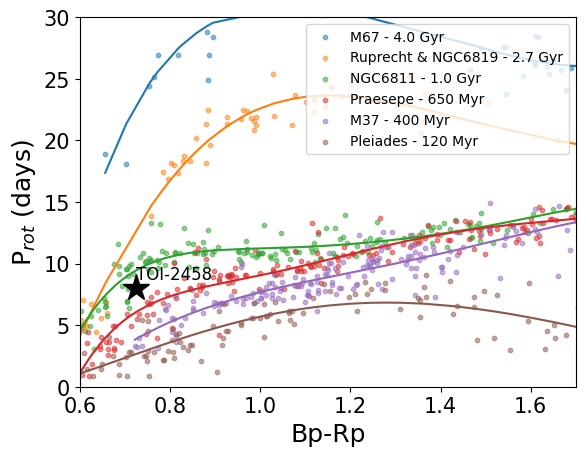}
\caption{Color-period diagram of TOI-2458 together with members of well-studied clusters. Lines represent the polynomial fit to each cluster sequence.} \label{fig:gyroch}
\end{figure}

To determine the age of TOI-2458, we compared its rotation period with the clusters analyzed in \citet{Godoy21}. These clusters include the Pleiades cluster ($\sim$120\,Myr), M37 cluster ($\sim$400\,Myr), the Praesepe cluster ($\sim$650\,Myr), NGC\,6811 ($\sim$1\,Gyr). Additionally, we used several older clusters: Ruprecht\,147 and NGC\,6819 clusters ($\sim$2.7\,Gyr) from \citet{Curtis20} and M\,67 cluster ($\sim$4\,Gyr) from \citet{Barnes16}. Using the dustmaps code \citep{Green18} and three-dimensional Bayestar dust maps \citep{Green19}, we estimated the reddening E(B-V) for each star in these clusters. We then employed the method from \citet{Gaia2018} to calculate reddening in the Gaia colors. Fig. \ref{fig:gyroch} shows the position of TOI-2458 in the rotation period vs Gaia DR3 G$_{Bp}$ minus G$_{Rp}$ magnitudes plane together with all studied clusters. Our findings show that the star TOI-2458, highlighted with a black star symbol, appears much younger than in other age-dating techniques. With an estimated age between $0.65-1.00$\,Gyr, TOI-2458 falls between the ages of the Praesepe cluster and NGC\,6811. It's worth noting that although the older clusters have few members with a measured rotation period at TOI-2458's color, the field stars in this area generally have longer rotation periods \citep{Dungee22}.

In contrast to gyrochronology, the techniques previously discussed for determining the age of TOI-2458 are consistent with the older ages derived from stellar isochrones. Furthermore, the predicted rotation period based on the $\log R'_{HK}$ index ranges from 23 to 40 days (longer than observed). However, the relationship between $\log R'_{HK}$ and the rotation period stems from a complex relationship that lacks complete theoretical understanding. In particular, the impact of increased stellar rotation resulting from planet engulfment on this correlation remains ambiguous. Similarly, the age derived from gyrochronology cannot be accurately ascertained for stars whose rotation is influenced by either a stellar or planetary companion \citep{Benbakoura19}. Therefore, we propose that recent planet engulfment may elucidate the atypical rotation period observed within the context of gyrochronology, as well as the relationship between $\log R'_{HK}$ and $P_{rot}$. Further elaboration on this concept is provided in Section \ref{sec:activity}, where we illustrate that similar unusual rotation periods are also observed for other stars comparable to TOI-2458.

\subsection{Frequency Analysis}\label{sec:Freq}

\begin{figure*}
\centering
\includegraphics[width=0.80\textwidth, trim= {0.0cm 0.0cm 0.0cm 0.0cm}]{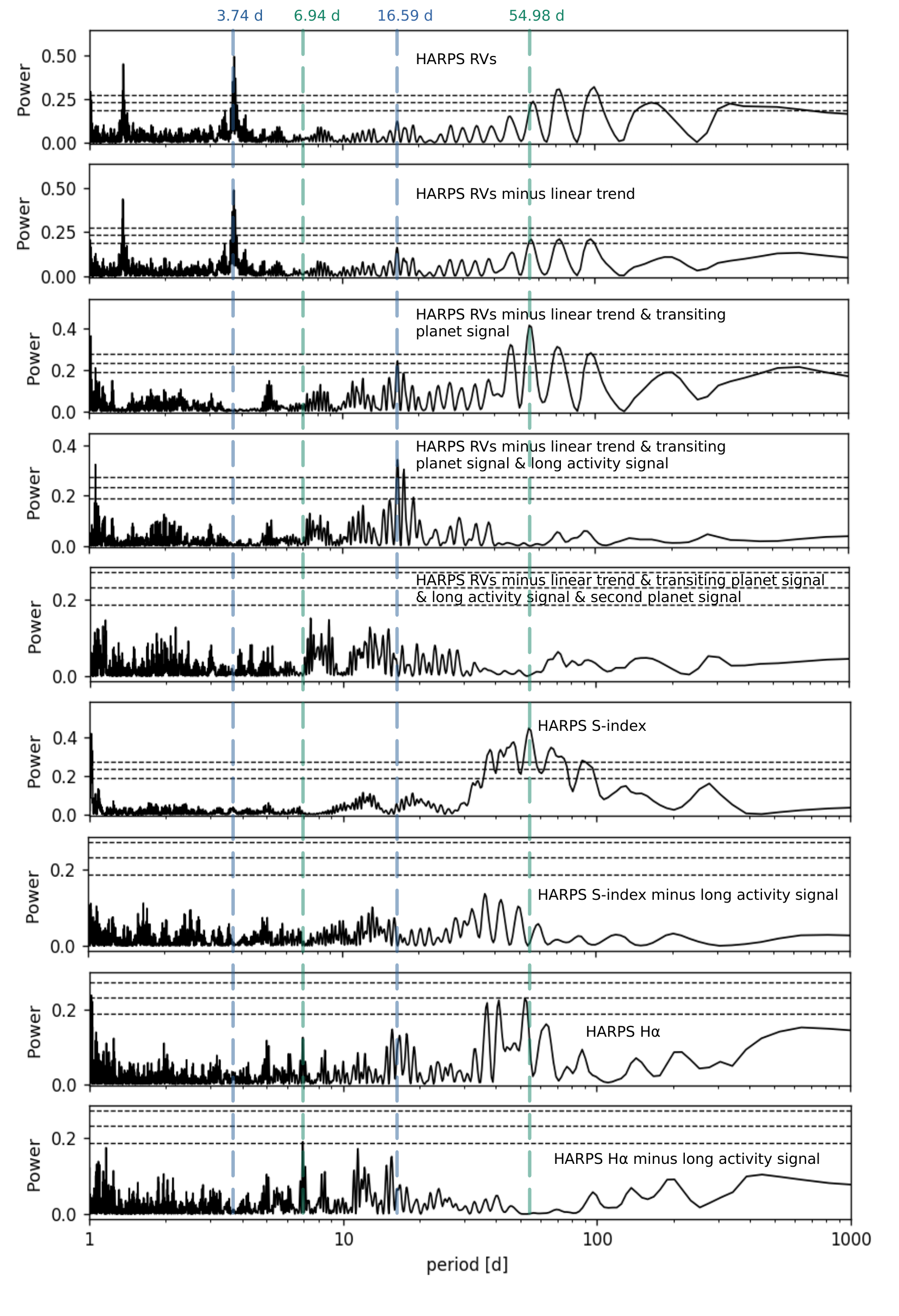}
\caption{Generalized Lomb-Scargle periodograms of HARPS RVs of TOI-2458 from top to bottom: (a) HARPS RVs, (b) HARPS RVs minus linear trend, (c) HARPS RVs minus linear trend and transiting planet signal, (d) HARPS RVs minus linear trend, transiting planet signal and long activity signal, (e) HARPS RVs minus linear trend, transiting planet signal, long activity signal and second planet signal, (f) HARPS S-index, (g) HARPS S-index minus long activity signal, (h) HARPS H$\alpha$, (i) HARPS H$\alpha$ minus long activity signal. The vertical blue lines highlight the orbital period of the transiting companion and the second companion. The vertical green lines highlight the stellar activity signals. The 6.94-day line indicates the peak observed in the H$\alpha$ activity indicator (last panel), which is near the rotation period identified in the TESS photometry. The 55-day line highlights the maximum signal from the S-index activity indicator. Horizontal dashed lines show the theoretical FAP levels of 10\%, 1\%, and 0.1\% for each panel.} \label{fig:per}
\end{figure*}

In an effort to distinguish the Doppler reflex motion from planetary candidates and stellar activity and to detect the existence of plausible additional signals, 
a frequency analysis was performed on the RVs, H$\alpha$, and S-index values of the spectra produced by the {\tt TERRA} pipeline. The GLS periodograms \citep{Zechmeister09} of the available time series were computed, and the theoretical false alarm probability (FAP) levels of 10\%, 1\%, and 0.1\% were determined, as depicted in Fig.~\ref{fig:per}. The observation baseline spans approximately 400\,d, resulting in a frequency resolution of roughly $1/400=0.0025$\,d$^{-1}$. Initially, we implemented the pre-whitening technique incorporating the potential linear trend. This method did not alter the signals identified in the periodogram, which are now more clearly delineated. The most significant signal observed in the periodogram of the RVs has a period of 3.74\,d, which is consistent with the period of the transiting planet from TESS photometry. After removing this signal, the periodogram of the RV residuals exhibits a signal with a period of $\sim$54\,d, which is also present in both activity indicators. Although several other peaks are visible due to observation sampling. We attribute this signal to a short-term magnetic cycle observed in a group of other F-type stars. We elaborate further on this matter in Section \ref{sec:activity}. Following the removal of the 54-d signal from RVs, an additional peak at 16.6\,d is observed, which we attempted to fit as the second planet in Section~\ref{sec:RVs}, as it does not have a counterpart in any activity indicator. It is noteworthy that a second peak near the 16.6-d peak of nearly equal amplitude is also observed. These two peaks are 1-year alias to each other, making it challenging to ascertain the real signal. In the joint modeling in Section \ref{sec:RVs}, a sufficiently wide prior was adopted to encompass both peaks.

In order to ensure the robustness of results obtained from the {\tt TERRA} pipeline, we conducted a sanity-check exercise using the {\tt SERVAL} pipeline \citep{Zechmeister2018} to extract RV measurements from the HARPS spectra. Our findings revealed that all signals reported earlier remained significant in both the {\tt TERRA} and {\tt SERVAL} datasets. 
Additionally, the 7-d signal was observed in the H$\alpha$ activity indicator after the removal of the long activity signal. While the 7-day signal in the H$\alpha$ activity indicator is at the 10\% FAP level, the presence of a peak roughly at the same periodicity detected in the TESS photometry (which we attribute to the rotation period of the star) supports the short-period scenario. The prominent peak in the S-index periodogram exhibits a period of approximately 55 days, with several additional peaks being visible, which are 1-year aliases of each other. It is important to note that periodograms of several other activity indicators from the {\tt TERRA} and {\tt SERVAL} pipelines did not reveal any of the discussed signals, and therefore, they are excluded from this study.

\begin{figure*}
\centering
\includegraphics[width=0.93\textwidth]{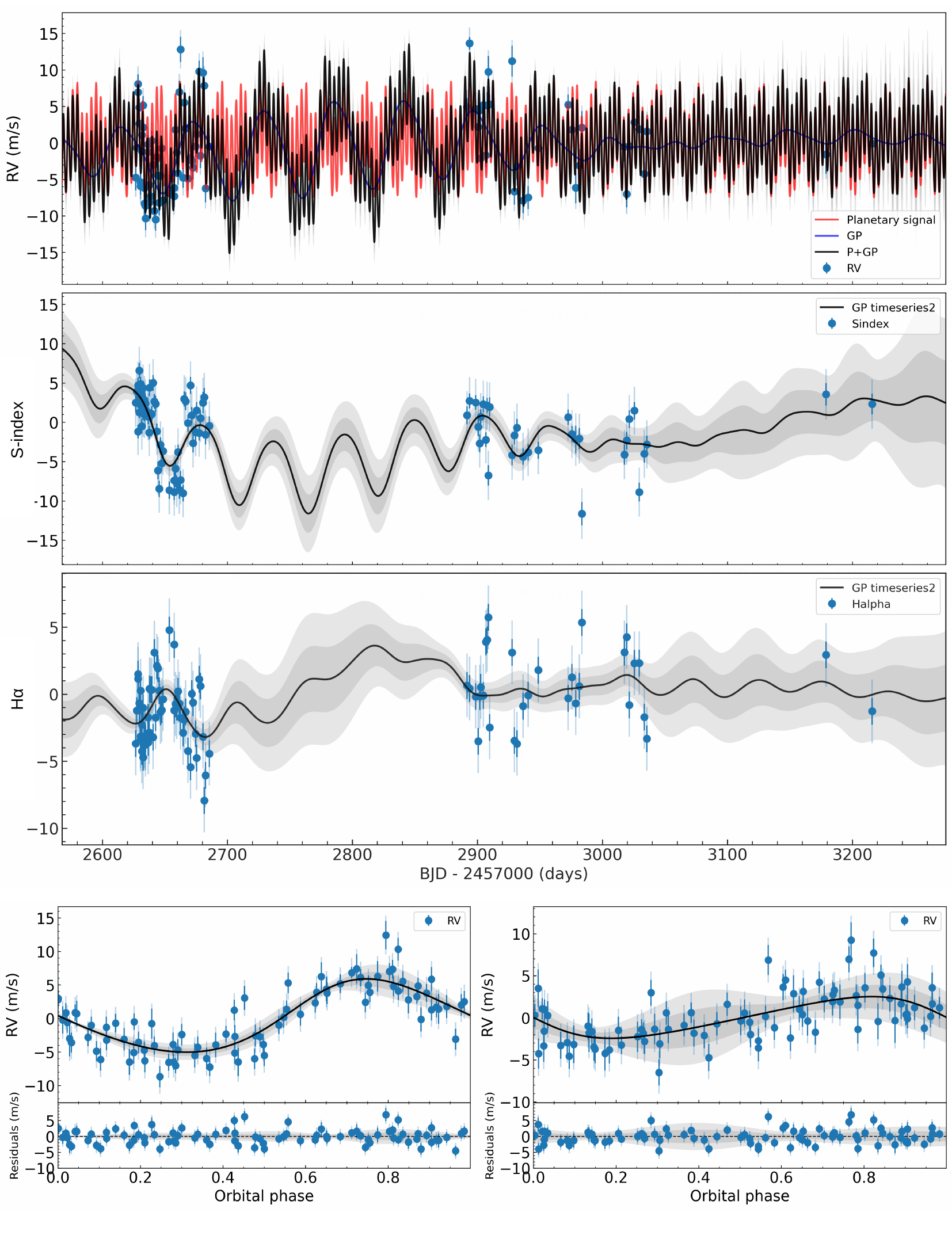}
\caption{The radial velocity curve of TOI-2458, fitted with the {\tt pyaneti} as described in Section \ref{sec:RVs}. The top panel shows the HARPS RV time series with all signals included. The middle panels show the HARPS S-index and H$\alpha$ time series, while the two bottom figures show the phased RV curves for TOI-2458\,b (left) and TOI-2458\,c (right). The middle panels present the outputs from two separate analyses using RVs with the S-index and RVs with H$\alpha$ respectively. This approach was used to save computational time, and we note that both analyses yield consistent results. The phased RV curves have been corrected for stellar activity using the GP model, which has been subtracted from the data.} \label{fig:RVs}
\end{figure*}

\begin{figure}
\centering
\includegraphics[width=0.46\textwidth, trim= {0.0cm 0.0cm 0.0cm 0.0cm}]{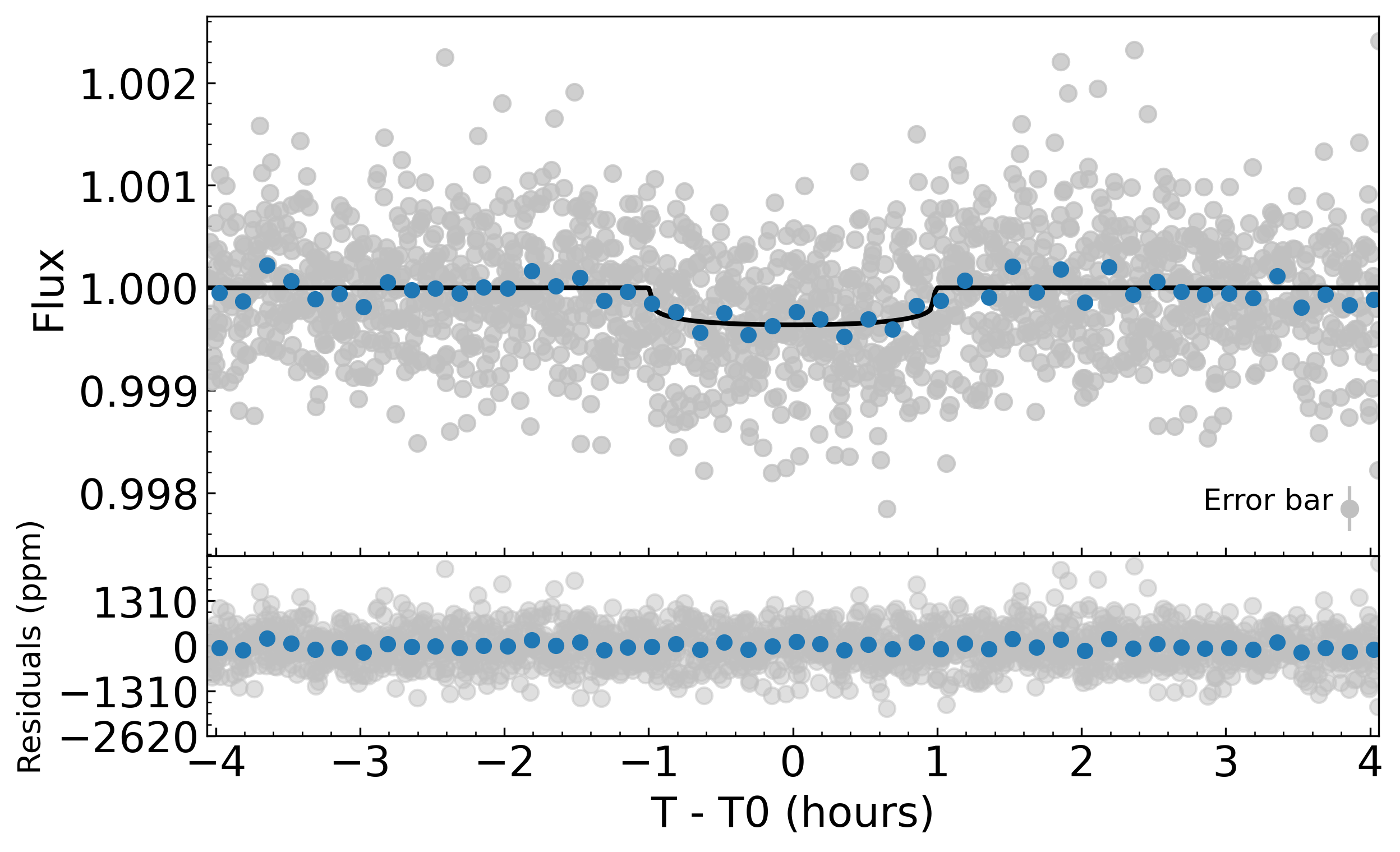}
\caption{The phased LC of TOI-2458\,b, fitted with {\tt pyaneti} as described in Section \ref{sec:RVs}. The grey points represent TESS data, and the blue points are TESS binned data in the phase curve with a bin width of 10 minutes in the phase. The black line represents the best transit and eclipse model.} \label{fig:transit}
\end{figure}

\subsection{Joint RVs and transits modelling}\label{sec:RVs}

We utilized the {\tt pyaneti} package \citep{pyaneti,Barragan22} to conduct a joint analysis of transit photometry and radial velocity data. {\tt Pyaneti} is a software program written in PYTHON/FORTRAN that utilizes a Markov chain Monte Carlo (MCMC) sampler based on EMCEE \citep{Foreman13} to generate marginalized posterior distributions of exoplanet RV and transit parameters. {\tt Pyaneti} is capable of conducting multiplanet fits with RV and transit data, together with multidimensional GP regression to simultaneously model stellar activity signals in Doppler data and activity indicators.

We have performed a comprehensive analysis of the transit photometry and radial velocity data. To account for the variability present in the TESS LC, we employed the activity-corrected LC that was processed using the procedure outlined in Section~\ref{sec:TESS}. As discussed in Section~\ref{sec:Freq}, the RV data contain stellar activity. Therefore, we incorporated a multidimensional Gaussian-process (GP) model, which has been previously described by \citet{Georgieva21} and \citet{Barragan22}, to account for the stellar variability. The RV data and the S-index activity indicator were modeled together, with the assumption that they could be described using the same underlying GP function. To achieve this, we utilized a quasi-periodic covariance function:

\begin{equation}\label{eq}
      {\gamma}_{QP,i,j} =
         { exp \left[ -\frac{sin^2\left[\pi(t_i-t_j)/P_{GP}\right]}{ 2{\lambda^2_P} } - \frac{(t_i-t_j)^2}{2{\lambda^2_e}}
                   \right]
}\,,
   \end{equation}

with P$_{GP}$ representing the period of the activity signal, $\lambda_p$ indicating the inverse of the harmonic complexity, and $\lambda_e$ representing the long-term evolution timescale. We have set the prior for $P_{GP}$ around the 54-day activity signal. The signal is clearly identifiable in RVs and activity indicators, and it corresponds to a short-term magnetic cycle observed in various F-type stars, as discussed in Section \ref{sec:activity}. It is important to note that when we repeated the analysis using the H$\alpha$ activity indicator instead of the S-index activity indicator, we obtained consistent results.

In this study, two models were compared, one with a single transiting planet and an activity, and the other additionally accounting for a second planet found in frequency analysis, as explained in Section \ref{sec:Freq}. The models were evaluated by computing the Bayesian model log evidence (ln\,Z), a widely used metric. If the difference in ln\,Z between the models is less than two, they are considered indistinguishable \citep{Trotta08}. The analysis produced $\triangle\,ln\,Z\,=\,16$, indicating a preference for the two-planet model. The quadratic limb darkening law was employed, with coefficients $q_1$ and $q_2$. The posteriors of fitted parameters and the physical parameters derived from them are presented in Table \ref{table:planet_par_pyan}. The RVs with the RV model, along with the phased RVs of both planets, are shown in Fig. \ref{fig:RVs}. The phased LC with the transit model is displayed in Fig. \ref{fig:transit}. Finally, the correlations between the main free parameters and the derived posterior probability distributions are plotted in Fig. \ref{fig:corr}. We wish to also highlight that we have tried remodeling the data, permitting a larger prior for the GP long-term evolution timescale of up to 1000\,days. Despite this adjustment, the fitted parameters for both planets remained consistent within their respective error bars.

Our study detects a Neptune-mass planet, with an RV semi-amplitude of $K\,=\,5.3\pm0.4$\,m\,s$^{-1}$, and a transit depth of $\delta\,=\,363\pm51$\,ppm. We derived an eccentricity of e\,=\,$0.085^{+0.072}_{-0.055}$. To investigate the significance of the non-circular orbit, we computed a Bayesian model log evidence by comparing models with fixed zero eccentricity and eccentricity as a free parameter. The resulting value of $\triangle\,ln\,Z\,=\,1.5$ indicates that we are unable to differentiate between a circular or non-circular orbit. 

The timing of the second planet's transit remains uncertain, and it could potentially be located anywhere within the TESS LC. Even the gap situated in the middle of the sector cannot be ruled out as a possible location. Despite the lack of any single-transit event indication in the TESS LC, given the planet's unknown size, it is still possible that the planet transits. Our analysis has revealed a minimum mass of approximately $10\,$M$_{\oplus}$, and the planet's orbital period is approximately 16.6\,d. The eccentricity of the planet remains unconstrained.

\subsection{Dynamical analysis}

The TOI-2458 system is fascinating from the point of view of its dynamical architecture and evolution. The orbit of the transiting planet is highly inclined to the stellar spin, which raises questions about how this rare configuration originated. While the semi-major axes of the planets are well determined, the inclination of the outer planet, TOI-2458\,c, is unknown. Currently, there is no way to fully determine this inclination and the mass of the outer planet except by analyzing the effects of the mutual gravitational interactions of the companions, which may constrain the orbital architecture of the system. Fortunately, our analysis shows that the long-term stability requirement and the expected orbital evolution within the physically allowed ranges are indeed meaningful in this case. We summarize our findings and provide all details of the analysis with corresponding figures in the Appendix \ref{dyn_a}.

We found that stable regions appear only in a portion of the parameter  ($\Delta\Omega,i_{c}$)-plane outside mutual inclinations approximately $i_{mut} \in (60^{\circ}, 140^{\circ})$. This implies moderate mutual inclinations. Considering that $i_{b} = 84^{\circ} \pm 0.5^{\circ}$, then $i_{c} \simeq (24^{\circ}, 156^{\circ})$ and its mass is in the range $M_{c} \simeq  (10, 25)$\,M$_{\oplus}$. The overall 3D dynamical structure of the TOI-2458 system is very complex despite apparently well-separated planets with small masses. The results of the dynamical simulations constrain the observed system to relatively small islands of moderate inner, close-in planet eccentricity when the orbital planes are moderately inclined.

\begin{figure*}
\centering
\includegraphics[width=0.80\textwidth]{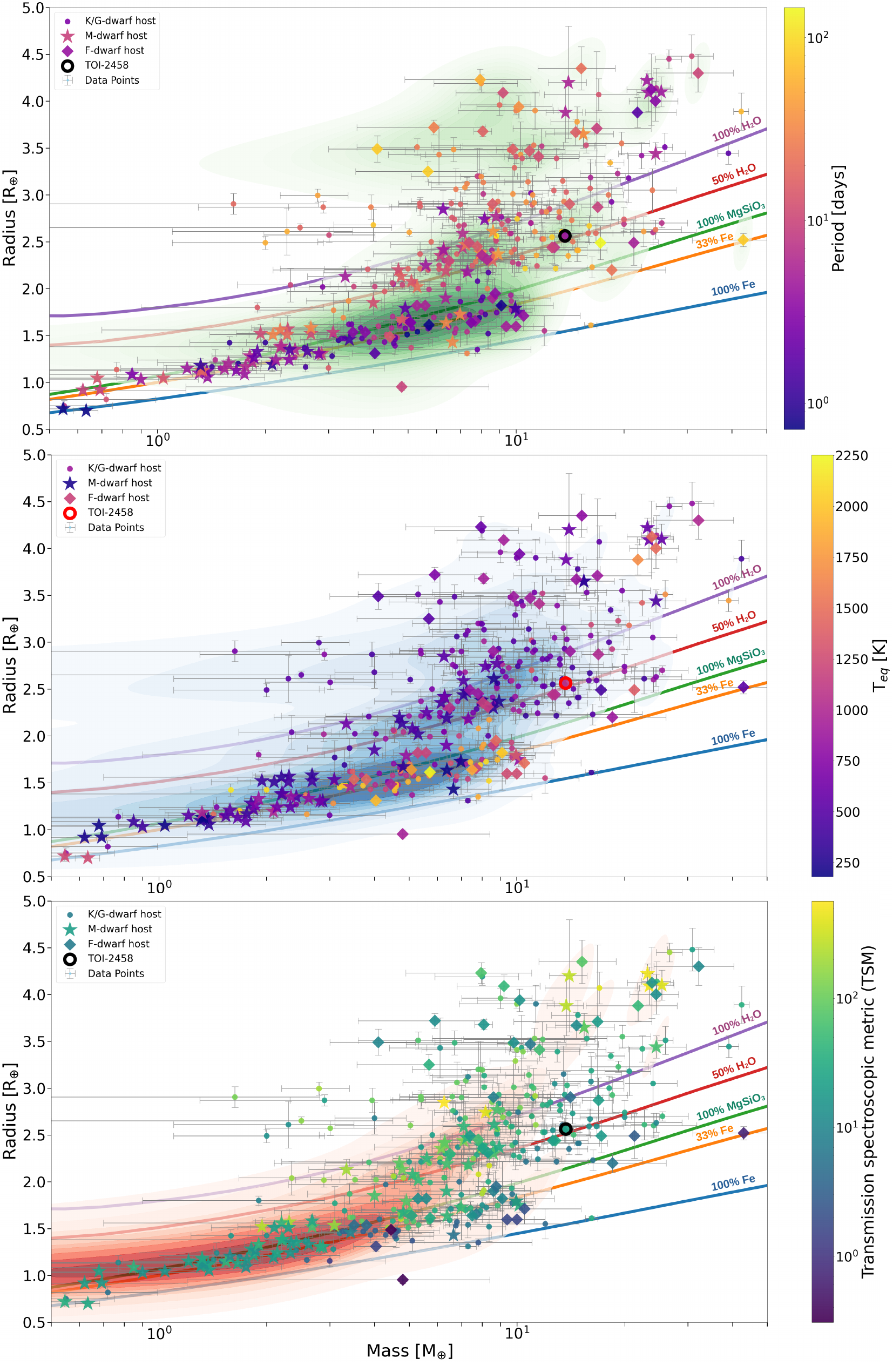}
\caption{The population of known companions within the mass interval $0.5-50 M_{\oplus}$. The position of TOI-2458\,b is highlighted. The top subplot is color-coded based on the orbital period of the planets, the middle one based on the equilibrium temperature, and the bottom one based on the transmission spectroscopic metric of the planets. The background shows the kernel density estimate (KDE) with green, blue, and red colors representing planets around F, K/G, and M stars, respectively. The mass-radius tracks from \citet{Zeng19} are utilized.} \label{fig:mas_radius}
\end{figure*}

\begin{figure*}
\centering
\includegraphics[width=1.0\textwidth]{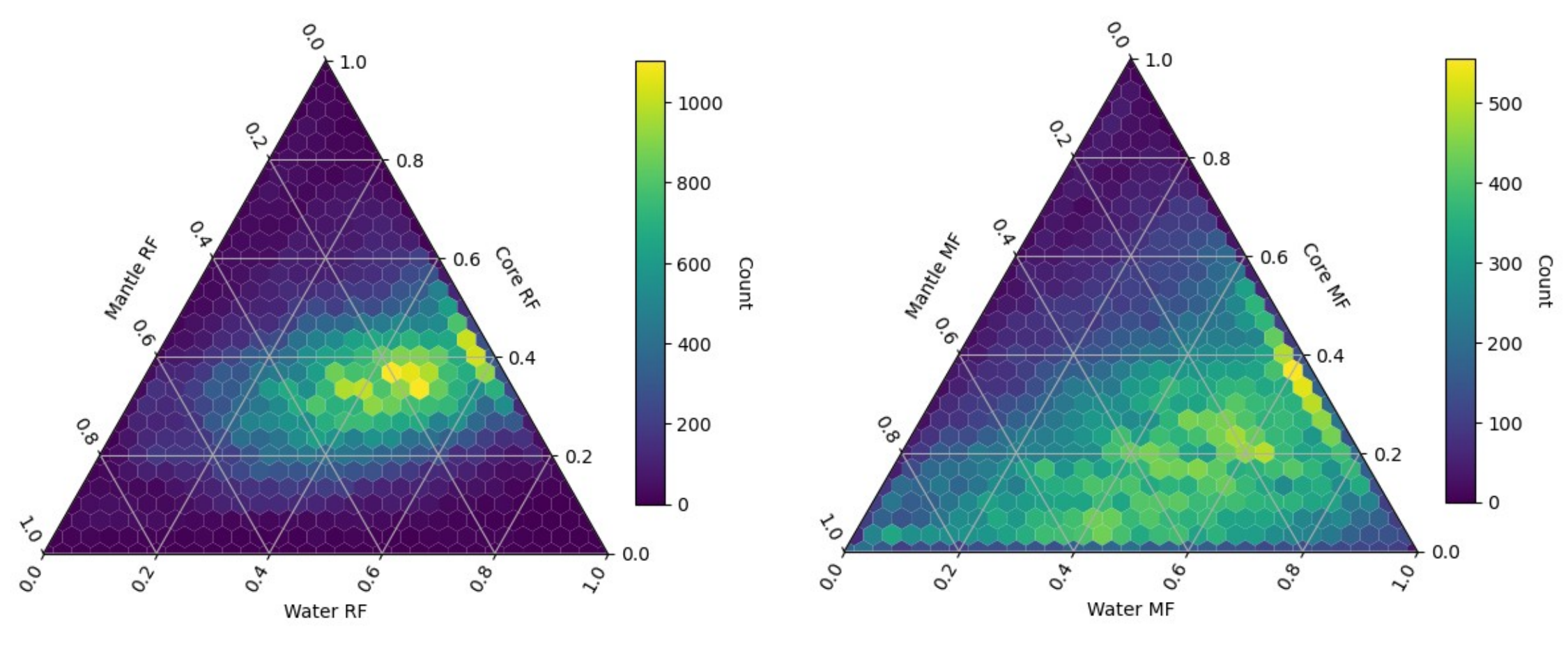}
\caption{Ternary plots of the interior structure of TOI-2458\,b. The left panel shows the radius fraction configuration and the right panel shows the mass fraction configuration.} \label{fig:interior}
\end{figure*}



%
%

\section{Discussion} \label{sec:discussion}
\subsection{Mass-radius diagram}\label{sec:mass-radius}

Combining data from the TESS mission and HARPS spectroscopy, we detected and confirmed the planetary nature of the P\,=\,3.7\,d transiting candidate around the F-type star TOI-2458. We found the fundamental parameters of TOI-2458\,b to be M$_p=13.31\pm0.99$\,M$_{\oplus}$ and R$_p=2.83\pm0.20$\,R$_{\oplus}$. Additionally, HARPS RVs revealed the presence of a second planet with a period of P\,=\,16.6\,d and a minimum mass of M$_p$sini\,=\,$10.22\pm1.90$\,M$_{\oplus}$.

We present a comparative analysis of the mass and radius of TOI-2458\,b with other well-characterized planets of comparable mass and size. Fig.~\ref{fig:mas_radius} showcases the analysis results through three subplots distinguished by their color coding based on different parameters such as orbital period, equilibrium temperature, and transmission spectroscopic metric. 
The placement of TOI-2458\,b on the mass-radius diagram is rather conventional, with several planets positioned within the middle of the radius range observed for planets orbiting F-type stars. These planets align with composition tracks corresponding to water content ranging from $50-100\%$ and exhibit moderate temperatures. Another subset of highly irradiated planets is situated near the composition track for pure silicate rocks. The last subset comprises typically long-period companions with radii of approximately 3.5\,R$_{\oplus}$. 
Additionally, due to the relatively large size of the primary star, the TOI-2458\,b's transmission spectroscopy metric of 25 is quite low, although comparable with other planets in this region. Nonetheless, it presents a challenging target for atmospheric characterization. When compared with the mass-radius tracks from \citet{Zeng19}, the planet appears to be composed of a large fraction of silicate rocks and water.

We can compare TOI-2458\,b with HD\,106315b \citep{Barros17}, as both exoplanets share comparable mass, radius, and stellar hosts. HD106315b, orbiting an F-type star with a period of 9.55 days, exhibits a mass of $12.6\pm3.2$\,M$_{\oplus}$ and a radius of $2.44\pm0.17$\,R$_{\oplus}$. Similarly to the TOI-2458 system, the HD106315 system features a second outer companion with a mass of $15.2\pm3.7$\,M$_{\oplus}$, a radius of $4.35\pm0.23$\,R$_{\oplus}$, and an orbital period of approximately 21 days. The authors discuss that the distinct densities of the two planets may stem from disparate formation regions.


TOI–2458\,b has a radius that positions it slightly above the known radius valley \citep{Fulton17}. There are currently several hypotheses behind the origin of these distinct populations of low-mass exoplanets, such as photo-evaporation \citep{Owen17}, core-powered mass loss \citep{Ginzburg18}, or migration after formation beyond the ice-line \citep{Mordasini09}.

\subsection{Transiting planet composition}\label{sec:comp}

We have utilized a machine learning model called {\tt ExoMDN} \citep{Baumeister23}, which is based on mixture density networks, to model the interior structure of TOI-2458\,b. This model has been trained on a massive dataset of over 5.6 million synthetic planets that are less than 25 Earth masses in order to create a highly accurate representation of their internal composition. This synthetic dataset, created using the {\tt TATOOINE} code \citep{Baumeister20,MacKenzie23}, consists of planets with an iron core, a silicate mantle, a water and a high-pressure ice layer, and a H/He atmosphere. {\tt ExoMDN} consists of two trained models: one trained on planetary mass, radius, and equilibrium temperature, and the second includes the fluid Love number k2 as an additional input, which is used to overcome degeneracy between layers. The fluid Love number k2 is a parameter that depends on the density distribution of the planet and can be computed for some planets, as highlighted in previous studies \citep{Csizmadia19,Akinsanmi19}. Another potential method to mitigate degeneracy in planetary interior modeling is to use the elemental abundances of the host star, which may provide insights into the planet's bulk composition and atmospheric composition \citep{Dorn15}. However, such an approach requires making additional assumptions regarding the planet's formation and evolutionary history. In the case of TOI-2458\,b, this planet does not conform to the known track of any planet with a k2 detection. Consequently, we have chosen not to incorporate such a constraint in our analysis, thereby broadening the range of potential interior configurations. To account for the uncertainties associated with planetary parameters, we model normal distributions with a standard deviation equivalent to the uncertainties of each parameter.

Analysis of TOI–2458\,b suggests a core-mass fraction of $0.20_{-0.18}^{+0.36}$, a mantle-mass fraction of $0.30_{-0.27}^{+0.47}$, a water-mass fraction of $0.43 \pm 0.36$, and a volatile element mass fraction of $0.001_{-0.001}^{+0.020}$. The core-radius fraction is estimated to be $0.29 \pm 0.16$, while the mantle-radius fraction is $0.22_{-0.19}^{+0.29}$, and the water-radius fraction is $0.34_{-0.26}^{+0.23}$. The atmosphere-radius fraction constitutes $0.12_{-0.09}^{+0.18}$ of the total radius. Due to the large uncertainty in these parameters, several different planetary configurations are possible, such as one where the mantle mass fraction is close to zero. To achieve a more precise determination of the radius, additional high-precision photometric observations are necessary. Unfortunately, it is very difficult to establish the actual structure of planets similar to TOI–2458\,b using mass and radius alone. These planets lie in a region of the parameter space where several composition tracks are in close proximity to each other. The mass-radius configuration of TOI-2458\,b is equally consistent with a water-rich world without an extended volatile atmosphere and a rocky core with an H$_2$ layer. To fully ascertain the nature of these scenarios, a further characterization of the planet's atmosphere is required. Ternary plots from the analysis are presented in Fig. \ref{fig:interior}.

\subsection{Stellar activity}\label{sec:activity}

We report on the activity cycle of TOI-2458, which exhibits a period of 53 days. This cycle is the second shortest on record for an F-type star. The first star discovered with a short cycle was $\tau$\,Boo (HD\,120136), an F7 spectral-type star with a hot Jupiter companion in a close orbit of 3.3\,days and an M2-type companion at the separation of 220\,AU. The rotation period of $\tau$\,Boo is found to be in close proximity to 3.3\,days, indicating synchronization. Nonetheless, this star is noteworthy because of additional variability in the Ca\,{\small{II}} lines, characterized by a surprisingly short period. This finding is supported by \citet{Baliunas97}, who determined a period of 116\,days, which was later confirmed by \citet{Mengel16} and \citet{Mittag17} investigating the S-index with the NARVAL and TIGRE instruments, respectively. \citet{Mittag17} further showed that the X-ray data support a periodicity of approximately 120\,days, raising an intriguing question about the representative nature of $\tau$\,Boo's fast activity period for main-sequence F stars. To address this question, \citet{Mittag19} employed an analysis of the S-index time series of F-type stars taken with the TIGRE telescope. They detected three additional short-term cycles and one candidate cycle between 0.5--1.0 year. Moreover, \citet{Subjak23} found that HD\,46588 is a twin of $\tau$\,Boo with respect to the activity cycle. The star is of the same F7 spectral type and hosts an L9 type companion at a projected physical separation of 1420\,AU \citep{Loutrel11}. Using the SONG spectrograph, the authors found evidence of a chromospheric cycle with a period of 127 days. Recently, \citet{Seach22} found that stars $\theta$\,Dra and $\beta$\,Vir also exhibit short-term activity cycles both well below 100 days.

We investigate a sample of stars exhibiting short-term activity cycles with the aim of identifying common characteristics among them, focusing on their age, rotation period, $R'_{HK}$ index, and close companions. The findings are reported in Table \ref{tab3}, which includes the relevant parameters for all stars. Our analysis reveals that TOI-2458 displays the second shortest cycle ever observed and is also the oldest system in the sample. However, we observe no discernible correlation between age and period. Our previous analysis (Section \ref{sec:age}) highlights that TOI-2458 exhibits a surprisingly short rotation period compared to its age. This may be a feature common to the sample, as, despite their advanced age, many of the stars rotate at a fast speed. Obvious are HD\,46588, $\theta$\,Dra, and $\beta$\,Vir, which, similarly to TOI-4987, exhibit low $R'_{HK}$ indexes and advanced ages yet also possess unexpectedly short rotation periods. $\beta$\,Vir is an analog to TOI-2458. The star has a rotation period of 9 days, an activity cycle of 83 days, an age of 3 Gyr, and a log\,$R'_{HK}$ of field stars (-4.91).

The rotational period of stars can be influenced by their nearby companions through tidal interactions \citep[e.g.,][]{Fleming19,Santos19}. Tidal friction causes the companion to migrate inwards and to spin up the star gradually. Furthermore, the presence of a hot Jupiter in close proximity to a star, as postulated by \citet{Fares09}, may be responsible for accelerating the star's cycle by synchronizing its outer convective envelope due to tidal interactions. As such, the presence of close companions can offer an explanation for observed phenomena. For instance, $\tau$\,Boo has a near companion with an orbital period of approximately 3.3\,days, which appears synchronized with the star's rotation period. However, most stars do not exhibit indications of having short-period giant companions. We cannot completely rule out this scenario, as their orbits' orientation may possibly hinder detection, or these planets may have been engulfed by their host stars. If the orbital angular momentum is too low to enable spin-orbit synchronization, tidal evolution would ultimately result in planet engulfment \citep{Hut80}. Massive planets on very tight orbits can tidally migrate inwards and spin up their star within the main-sequence lifetime \citep{Carone07,Jackson09,Tejada21}. In their study, \cite{Tejada21} have uncovered evidence suggesting that hot Jupiters modify the rotation of their host stars during the main sequence. This determination was made by comparing the rotation periods of Sun-like stars hosting hot and cold Jupiters, as well as low-mass planets. Moreover, \cite{Jackson09} have presented evidence indicating that the observed orbital distribution of exoplanets is consistent with the predictions of tidal theory. This suggests that the tidal destruction of close-in exoplanets is a common phenomenon. For instance, the planetary system WASP-12 is notable for hosting a hot Jupiter with a decaying orbit \citep{Yee20}.

\begin{figure}
\centering
\includegraphics[width=0.46\textwidth, trim= {0.0cm 0.0cm 0.0cm 0.0cm}]{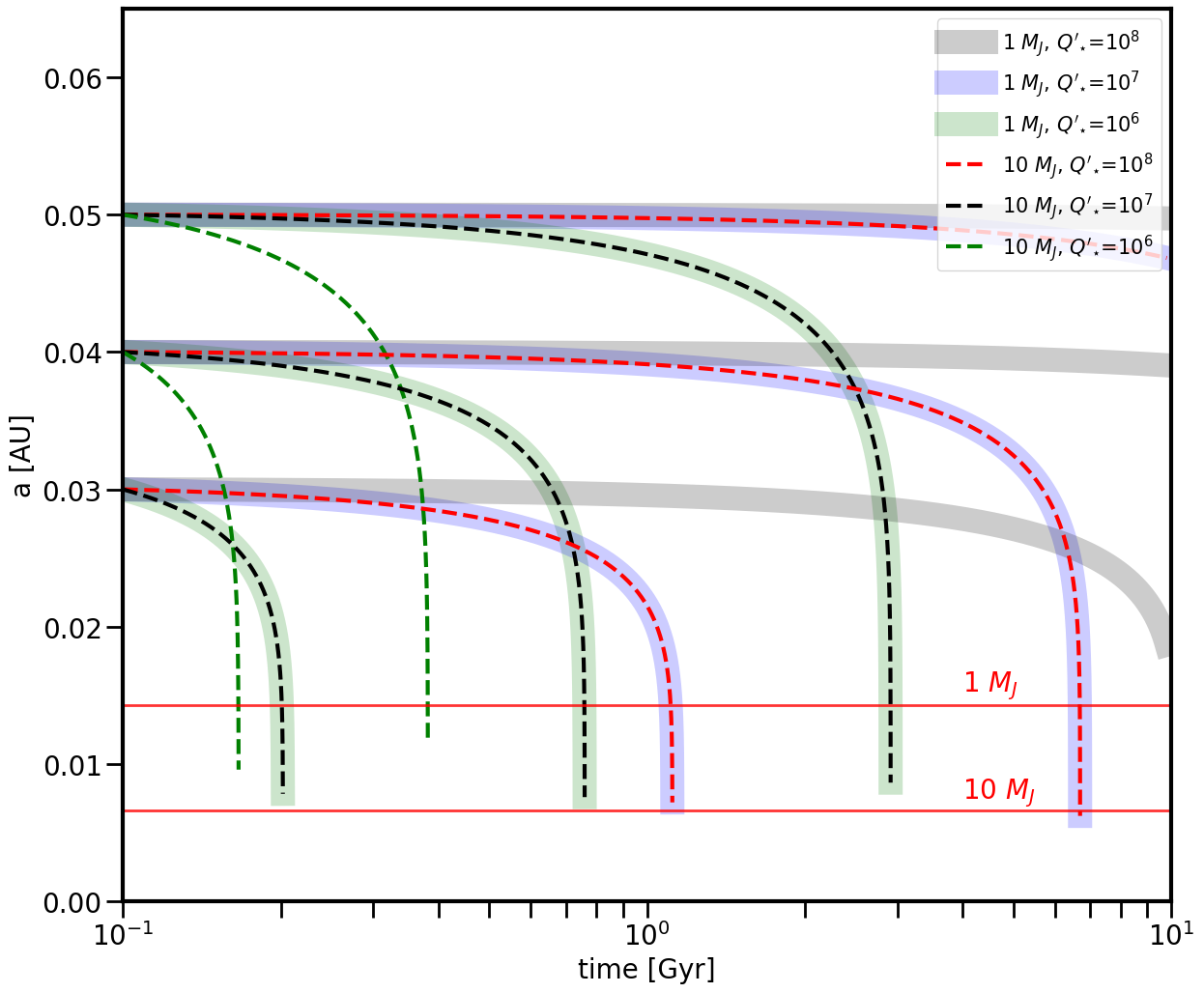}
\caption{Tidal evolution of the orbit of TOI-2458\,b. The horizontal red lines indicate the Roche limit for both planet scenarios.} \label{fig:tidals}
\end{figure}

We employ the same equations used by \citet{Jackson09} to investigate the tidal evolution of planets with masses of 1\,Jupiter mass and 10\,Jupiter masses orbiting F-type stars. Specifically, we examine the orbital decay of these planets and the time required for them to cross the Roche limit around the host star, beyond which they would be tidally disrupted and accreted by the star. The equations we use assume that the period of rotation for host stars is always substantially longer than the planet's orbital period. We also assume zero eccentricity, meaning that only the tides raised on the star by the planet contribute to its tidal evolution. Figure \ref{fig:tidals} shows a plot of the tidal evolution of the semimajor axis. The results indicate that the time taken by planets to reach the Roche limit is often less than 1\,Gyr for several initial conditions and decreases for higher-mass companions at smaller orbital distances. \citet{Oetjens20,Ahuir21} studied the influence of planetary engulfment on the stellar rotation of main-sequence stars. Their results indicate a large increase in stellar rotation velocity while the star is still in the main-sequence evolutionary phase. It is worth noting that such an over-rotation is likely to last for a few billion years \citep{Ahuir21}. In the subsequent section, we explore the possibility of an unobserved hot Jupiter companion in the TOI-2458 system.

\begin{table*}
	\centering
	\caption{Sample of F-type stars with a short activity cycle. P$_{cyc}$ is a period of a short activity cycle.}
	\label{tab3}
	\scalebox{0.82}{
	\begin{tabular}{lcccccccccr} 
		\hline
		\hline
		System       & TOI-2458 & $\tau$Boo & HD\,46588 & HD\,16673 & HD\,49933 & HD\,100563 & HD\,75332 & $\theta$\,Dra & $\beta$\,Vir\\
\hline \\
Age (Gyr) & $5.7_{-0.8}^{+0.9}$ & $0.85_{-0.41}^{+0.81}$ & $2.1_{-0.8}^{+1.3}$ & $1.20_{-0.55}^{+1.20}$ & $2.2_{-0.3}^{+0.3}$ & $1.4_{-0.4}^{+0.6}$ & $2.0_{-0.5}^{+0.6}$ & $2.1_{-0.2}^{+0.2}$ & $3.1_{-0.8}^{+0.8}$ \\
P$_{Rot}$ (days) & $8.9_{-2.5}^{+3.9}$ & $3.3 \pm 0.4$ & $5.1 \pm 0.6$ & $6.0 \pm 0.8$ & $3.5 \pm 0.3$ & $2.0 \pm 0.3$ & $3.75 \pm 0.35$ & $2.9 \pm 0.1$ & $9.2 \pm 0.3$ \\
log $R'_{HK}$ & $-4.98 \pm 0.06$ & $-4.67 \pm 0.06$ & $-4.88 \pm 0.06$ & $-4.60 \pm 0.06$ & $-4.50 \pm 0.06$ & $-4.64 \pm 0.06$ & $-4.46 \pm 0.06$ & $-4.68 \pm 0.06$ & $-4.91 \pm 0.06$ \\
P$_{cyc}$ (days) & 53 & 120 & 127 & 309 & 212 & 223 & 180 & 43 & 83\\
		\hline
		\hline \\
	\end{tabular}
	}
\end{table*}

\subsection{System architecture}\label{sec:composition}

The TOI-2458 system is a promising candidate for testing the hypothesis of the presence of hot Jupiters in systems showing short activity cycles. The system's transiting small planet that likely orbits close to the star's rotational poles provides an opportunity to investigate whether such an orientation is dynamically consistent with such an idea. The abundance of hot, low-mass planets with a mass of less than $30~M_{\oplus}$ on close-in orbits with a period of fewer than 100 days \citep[e.g.,][]{Mulders15,Winn15}, suggests that many in situ formed hot Jupiters conglomerate within systems that also contain low-mass planets. \citet{Batygin16} have studied the in situ formation and dynamical evolution of such hot Jupiter systems and demonstrated that under conditions appropriate to the inner regions of protoplanetary disks, rapid gas accretion can be initiated by 10--20 Earth-mass planets, which are very common. Their findings suggest that the interaction of in-situ formed hot Jupiters with outer low-mass planets has observable consequences. They determined a critical semimajor axis of low-mass planets beyond which secular resonant excitation of mutual inclination of both planets is guaranteed. Therefore, in the case of an almost circular orbit of a hot Jupiter, dynamical interactions during the early stages of planetary systems' lifetimes should highly increase the mutual inclinations for the specific range of the semimajor axis. Disk lifetime, stellar rotation rate, and semimajor axis of hot Jupiter are among the critical parameters that influence this specific range. Figure~9 from \citet{Batygin16} indicates that TOI-2458\,b can be located within this range in a large portion of parameter space, which could explain the observed low inclination of its orbit.

\begin{figure}
\centering
\includegraphics[width=0.48\textwidth, trim= {0.0cm 0.0cm 0.0cm 0.0cm}]{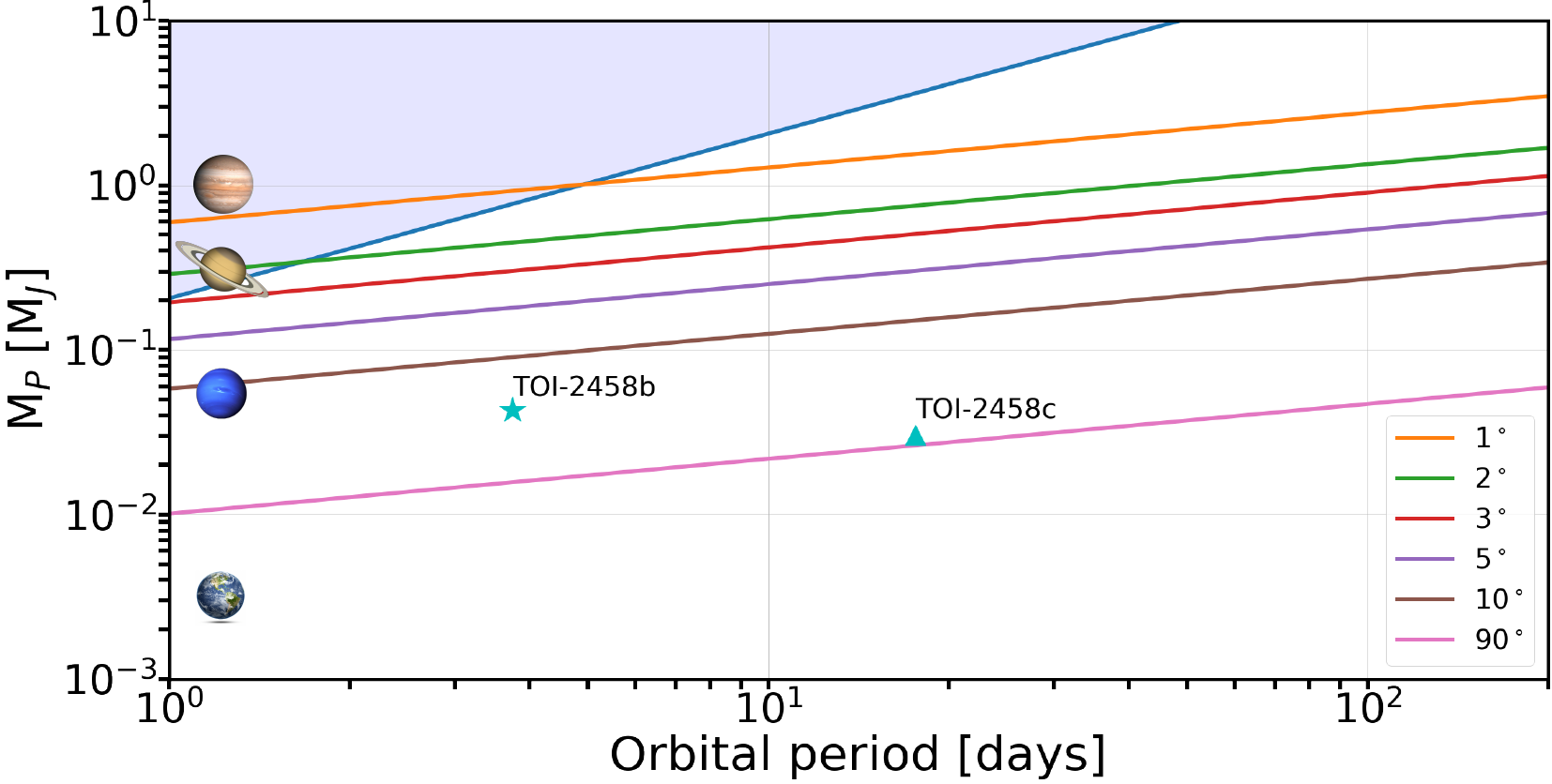}
\caption{Detection limits for companions around TOI-2458 based on their orbital inclination. The blue line represents the boundary for planets that would cause a strong tidal spin-up of TOI-2458.} \label{fig:limits}
\end{figure}

Not only is the TOI-2458 system consistent with the scenario of in situ formed hot Jupiter, but the orbital orientation of TOI-2458\,b would be its natural consequence. If the hot Jupiter has not yet been engulfed, its detection in RVs and photometry would be prevented due to its orbital orientation. Based on our radial velocity observations, we have determined the minimum orbital inclination required to prevent the detection of a hot Jupiter close and massive enough to affect the tidal spin-up of TOI-2458. We used the criterium from \cite{Tejada21} to define the theoretical mass and period boundary for planets causing strong tidal spin-up. \cite{Tejada21} showed that objects within this boundary spin up their hosts, which tend to spin faster than the stars hosting cold Jupiters. Our findings suggest that the orbital inclination would need to be below two degrees to prevent the detection of a hot Jupiter with a period of three days (see Fig. \ref{fig:limits}), quite an improbable scenario indicating that any hot Jupiter present may have been ingested. To date, no planet system has been observed that convincingly satisfies the implications described in \citet{Batygin16}. WASP-47 may be an undisrupted system with an in situ formed hot Jupiter, where an increase of mutual inclination has been avoided. In this context, TOI-2458 represents evidence supporting the in situ formation of a hot Jupiter in this system, with consequent dynamical evolution matching the observed orbital orientation of TOI-2458\,b. Due to the low $v \sin{i}$, it is not possible to measure the misalignment of TOI-2458\,b's orbit using the Rossiter-McLaughlin effect (see Table \ref{table:planet_par_pyan}). One more time, we emphasize the importance of confirming the stellar rotation period to thoroughly validate the system architecture. 

\subsection{Potential debris disc}\label{sec:debris}

TOI-2458 shows an infrared excess at 22 $\mu$m in the WISE/W4 bandpass,
with an excess level of 75 $\%$ (4.7 $\sigma$) compared to its
photospheric flux \citep{cruzsaenz14}. High-resolution imaging confirms the absence of photometric contaminants around the star, suggesting that the infrared excess may be attributed to a debris disc. While debris discs are typically associated with stars younger than 1\,Gyr \citep[e.g.,][]{siegler07, gaspar09, mizuki24}, it's important to note that the presence of such a disc does not necessarily indicate the youth of the system. Indeed, some older systems have been found to have debris discs, although their occurrence is less common compared to younger stars \citep[e.g.,][]{Trilling08}.

Small dust grains present in debris discs are the result of a collisional cascade, where planetesimals are gradually reduced in size \citep{wyatt08}. The timescale for the crossing of orbits within a planetesimal belt due to planet stirring can be calculated using equation (15) from \cite{mustillwyatt09}, taking into consideration the characteristics of both the planet and the disc. The disc temperature of $166_{-30}^{+42}$\,K from \cite{cruzsaenz14} was employed to determine the location of the debris disc at $4.0_{-1.5}^{+2.2}$\,AU using the equation (1) in \cite{Trilling08}. These values were used to illustrate the planetary perturbation for the disc in the TOI-2458 system. The timescale of secular perturbation for planet TOI-2458\,b varies from 0.2\,Gyr to 14\,Gyr, depending on the precise position of the disc. Planet TOI-2458\,c, which orbits slightly beyond the transiting planet, exhibits a larger orbital eccentricity uncertainty and only a lower limit for its mass, indicating its potential role as a perturber for the debris disc. Nevertheless, we note that unseen outer planets in proximity to the debris disc may also exert a more substantial influence on the belt's dynamics. To comprehensively understand the architecture of the TOI-2458 planetary system, encompassing both planets and a disc, future research endeavors could concentrate on confirming the debris disc and characterizing its location through spectroscopic observations.

\section{Summary} \label{sec:summary}

We have presented a discovery of the transiting mini-Neptune TOI-2458\,b around an F-type star. TOI-2458\,b has an orbital period of P\,=\,$3.73659\pm0.00047$\,days, a mass of $M_p=13.31\pm0.99\,M_{\oplus}$ and a radius of $R_p=2.83\pm0.20\,R_{\oplus}$. The star appears to be relatively old, with an estimated age of approximately 5.7\,Gyr. Given this age, we found an unexpectedly fast stellar rotation period of $8.9_{-2.5}^{+3.9}$\,days. Using a combination of $P_\mathrm{rot}$, $v \sin{i}$, and $R_\star$, we constrained a low stellar inclination angle of $i_\star\,=\,10.6_{-10.6}^{+13.3}$\,degrees. The star also shows a short-term activity cycle of 53\,days, which contributes to the increasing sample of F-type stars exhibiting similar behavior. Our investigation of a sample of stars exhibiting short-term activity cycles revealed that a short stellar rotation period may be a common characteristic. Such a rotation period may be explained by tidal interactions with hot Jupiters, which may be engulfed on timescales of a few Gyr or less. The TOI-2458 system appears to be consistent with the scenario of an in situ formed hot Jupiter, which may explain the stellar properties (rotation period and magnetic cycle) and orientation of the transiting companion TOI-2458\,b. Finally, the RV periodogram revealed the presence of a second planet in the HARPS data with a period of P\,=\,$16.55\pm0.06$\,days and a minimum mass of $M_p \sin i=10.22\pm1.90$~$M_{\oplus}$. Using dynamical stability analysis, we constrained the mass of this planet to the range $M_{c} \simeq  (10, 25)$\,M$_{\oplus}$.

Overall, our findings suggest that TOI-2458 is an interesting system for further study, particularly in the context of hot Jupiter formation and its influence on short-term activity cycles of F-type stars.

%
%
\begin{acknowledgements}
J.\v{S}. would like to acknowledge the support from GACR grant 23-06384O.
C.M.P. gratefully acknowledges the support of the  Swedish National Space Agency (DNR 65/19 and  177/19).
D.J. and G.N. gratefully acknowledge the Centre of Informatics Tricity Academic Supercomputer and networK (CI TASK, Gda\'nsk, Poland) for computing resources (grant no. PT01016).
G.N. thanks for the research funding from the Ministry of Science and Higher Education programme the "Excellence Initiative - Research University" conducted at the Centre of Excellence in Astrophysics and Astrochemistry of the Nicolaus Copernicus University in Toru\'n, Poland.
This work is partly supported by JSPS KAKENHI Grant Number
JPJP24H00017 and JSPS Bilateral Program Number JPJSBP120249910.
J.K. acknowledges the Swedish Research Council (VR: Etableringsbidrag 2017-04945).
S.M.\ acknowledges support by the Spanish Ministry of Science and Innovation through AEI under the Severo Ochoa Centres of Excellence Programme 2020--2023 (CEX2019-000920-S).
P.K. is grateful for the support from grant LTT-20015.
HJD acknowledges support from the Spanish Research Agency of the Ministry of Science and Innovation (AEI-MICINN) under grant PID2019-107061GB-C66, DOI:
10.13039/501100011033.
R.A.G. acknowledges the support from the PLATO Centre National D'{\'{E}}tudes Spatiales grant.
DR was supported by NASA under award number NNA16BD14C for NASA Academic Mission Services.
KAC acknowledges support from the TESS mission via subaward s3449 from MIT.
TGW acknowledges support from the UKSA and the University of Warwick.

Funding for the TESS mission is provided by NASA's Science Mission Directorate.

We acknowledge the use of public TESS data from pipelines at the TESS Science Office and at the TESS Science Processing Operations Center.

Resources supporting this work were provided by the NASA High-End Computing (HEC) Program through the NASA Advanced Supercomputing (NAS) Division at Ames Research Center for the production of the SPOC data products.


This paper includes data collected by the TESS mission that are publicly available from the Mikulski Archive for Space Telescopes (MAST).


This research has made use of the Exoplanet Follow-up Observation Program (ExoFOP; DOI: 10.26134/ExoFOP5) website, which is operated by the California Institute of Technology, under contract with the National Aeronautics and Space Administration under the Exoplanet Exploration Program.

The PLATOSpec team would like to thank observers Marek Skarka, Ji\v{r}\'{i} Srba, Zuzana Balk\'{o}v\'{a}, Petr \v{S}koda and Lud\v{e}k \v{R}ezba.

This work makes use of observations from the LCOGT network. Part of the LCOGT telescope time was granted by NOIRLab through the Mid-Scale Innovations Program (MSIP). MSIP is funded by NSF.

This publication was produced within the framework of institutional support for the development of the research organization of Masaryk University.

This research has made use of the Simbad and Vizier databases, operated at the centre de Donn\'ees Astronomiques de Strasbourg (CDS), and of NASA's Astrophysics Data System Bibliographic Services (ADS).

This work has made use of data from the European Space Agency (ESA) mission {\it Gaia} (\url{https://www.cosmos.esa.int/gaia}), processed by the {\it Gaia} Data Processing and Analysis Consortium (DPAC, \url{https://www.cosmos.esa.int/web/gaia/dpac/consortium}). Funding for the DPAC
has been provided by national institutions, in particular, the institutions participating in the {\it Gaia} Multilateral Agreement.

This work made use of \texttt{TESS-cont} (\url{https://github.com/castro-gzlz/TESS-cont}), which also made use of \texttt{tpfplotter} \citep{Aller20} and \texttt{TESS-PRF} \citep{Bell22}.
\end{acknowledgements}

\bibliographystyle{aa}
\bibliography{astro_citations}
\twocolumn
\appendix
\renewcommand\thefigure{\thesection.\arabic{figure}}    
\section{Dynamical analysis}\label{dyn_a}

First, to investigate the possible proximity of the TOI-2458 system to mean-motion resonances (MMRs) and the dynamical stability of our best-fit solution, we used direct $N$-body integrations and a CPU-efficient fast indicator called the Reversibility Error Method \citep[REM;][]{Panichi_Gozdziewski_Turchetti-2017MNRAS.468..469P}. REM is closely related to the Maximum Lyapunov Exponent (MLE). It is based on integrating the equations of motion with a time-reversible (symplectic) scheme forward and backward in time for the same number of time steps. Then, the difference between the initial and final states of the system, normalized by the size of the phase-space trajectory, makes it possible to distinguish between regular and chaotic evolution. REM is scaled so that $\log\mathrm{\widehat{REM}}=0$ means that the difference between the initial and final states is of the order of the orbit size.

\begin{figure}[!h]
\centering
\includegraphics[width=0.925\linewidth]{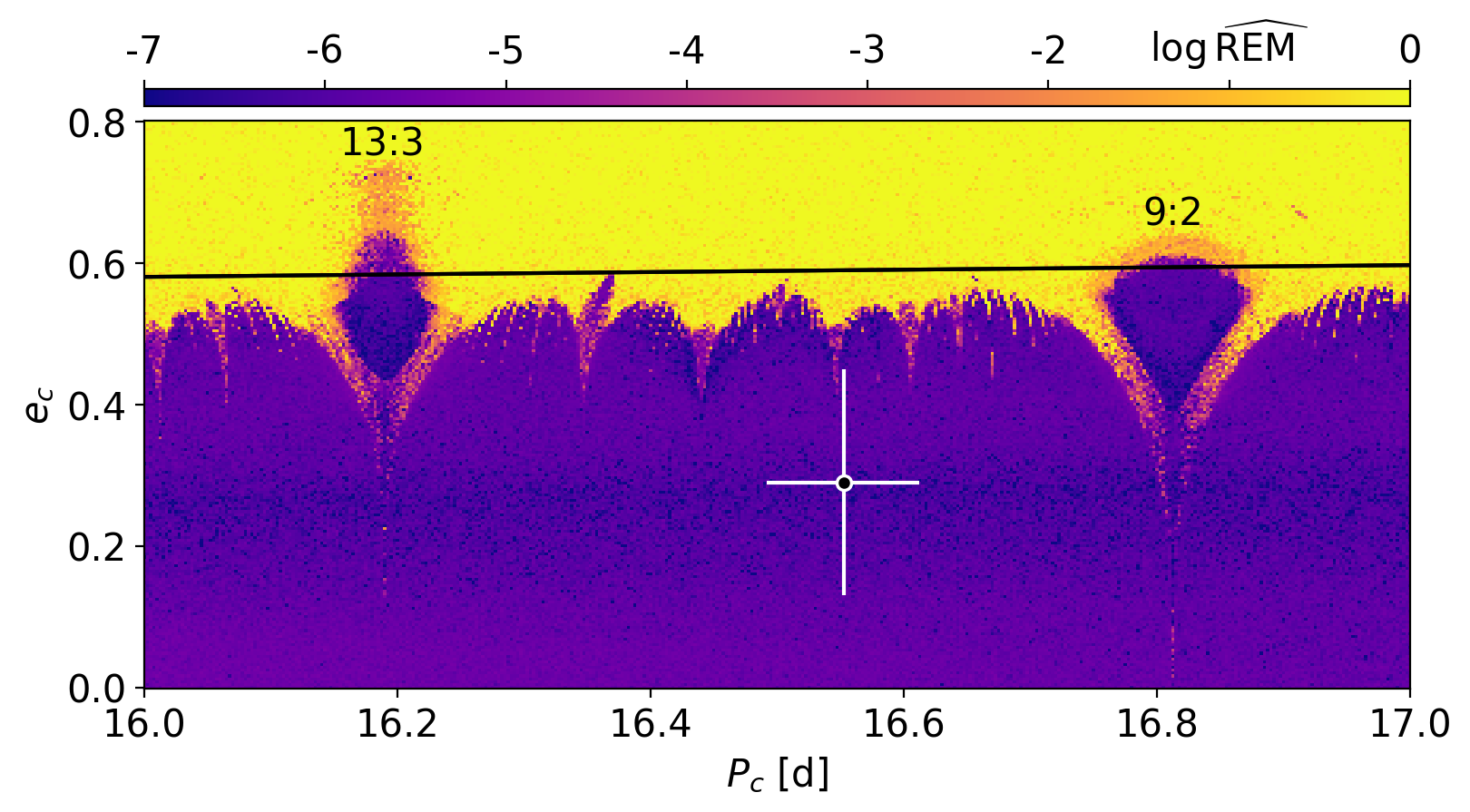}
\caption{
REM dynamical map for the solution shown in Table \ref{table:planet_par_pyan} in the $(P_{c},e_{c})$-plane of the planet c, the system is co-planar. Small $\log \mathrm{\widehat{REM}}$ characterize regular (long-term stable) solutions, marked with black/dark blue color. Dynamically chaotic solutions are marked with lighter colors up to yellow. The black curve is for the geometric collision of the orbits, defined by the condition: $a_{\rm b}(1 + e_{\rm b}) = a_{\rm c}(1 - e_{\rm c})$. The resolution of the map is 401~$\times$~201 points.}
\label{fig:mmrmap}
\end{figure}

We integrated REM with the \whfast{} integrator with the 17th order corrector, with a fixed time step of 0.05\,days, as implemented in the \rebound{} package \citep{Rein_Liu-2012A&A...537A.128R,Rein_Spiegel-2015MNRAS.446.1424R} for $\simeq 2\times 10^5$ orbital periods of the outer planet. As shown in Fig.~\ref{fig:mmrmap}, $\log\mathrm{\widehat{REM}}$ $\leq$ -5, indicating a stable solution. Furthermore, the location of the best-fitting model in phase space indicates its safe, non-resonant character. The orbital period of TOI-2458\,c was determined to be $P_{c}= 16.553^{+0.059}_{-0.061}${}d, and the dynamical map shows the nominal system shifted by a few $\sigma$ from nearby strongest 13:3 MMR and 9:2 MMR, respectively. It is worth noting that the eccentricity error is quite large, which may indicate proximity to the collision curve and increased mutual interactions between the planets.

\begin{figure*}
\centering
\hbox{
\includegraphics[width=0.333\linewidth]{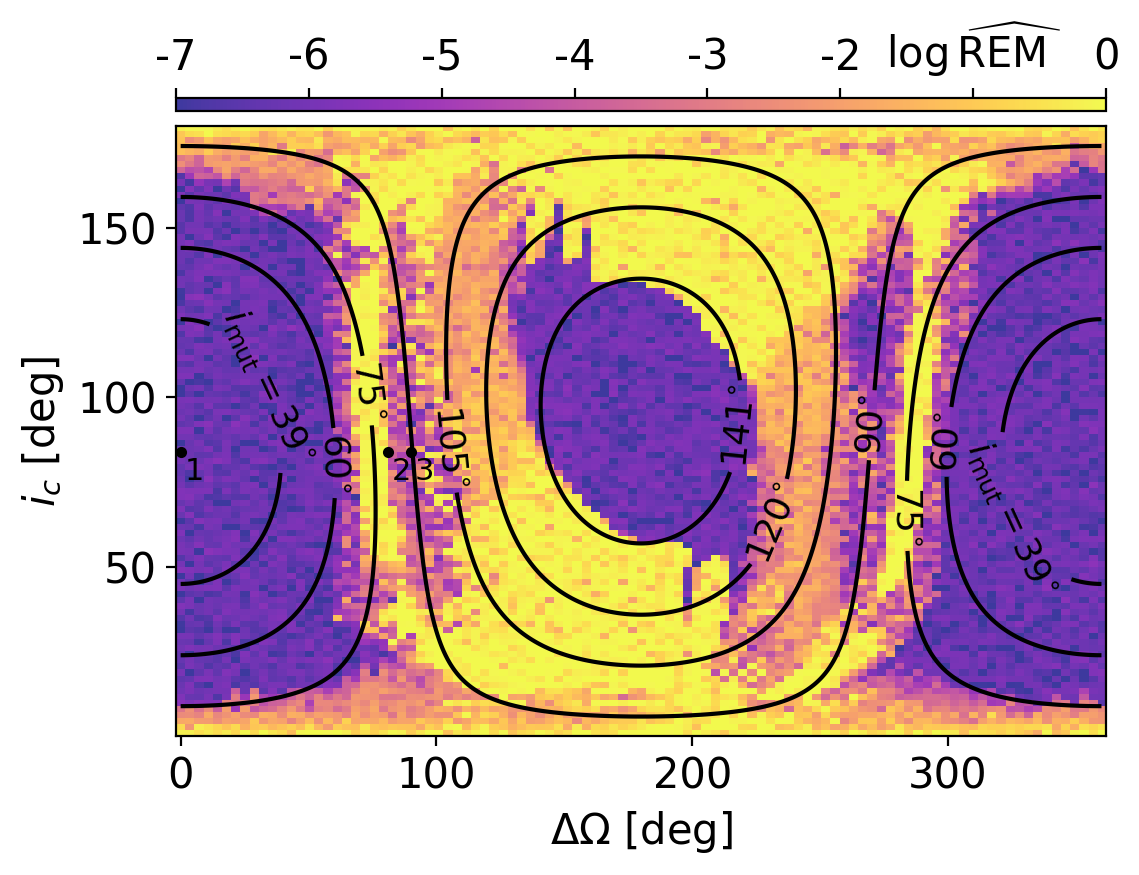}
\includegraphics[width=0.333\linewidth]{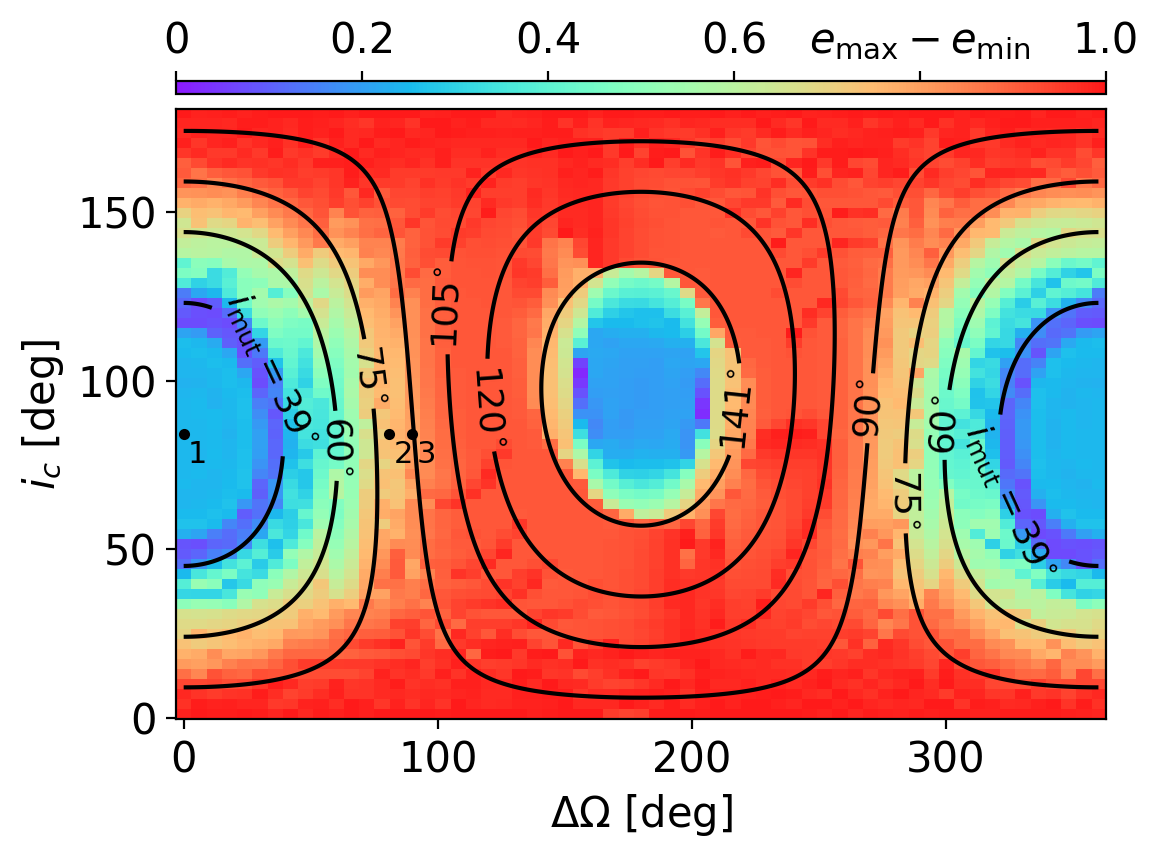}
\includegraphics[width=0.333\linewidth]{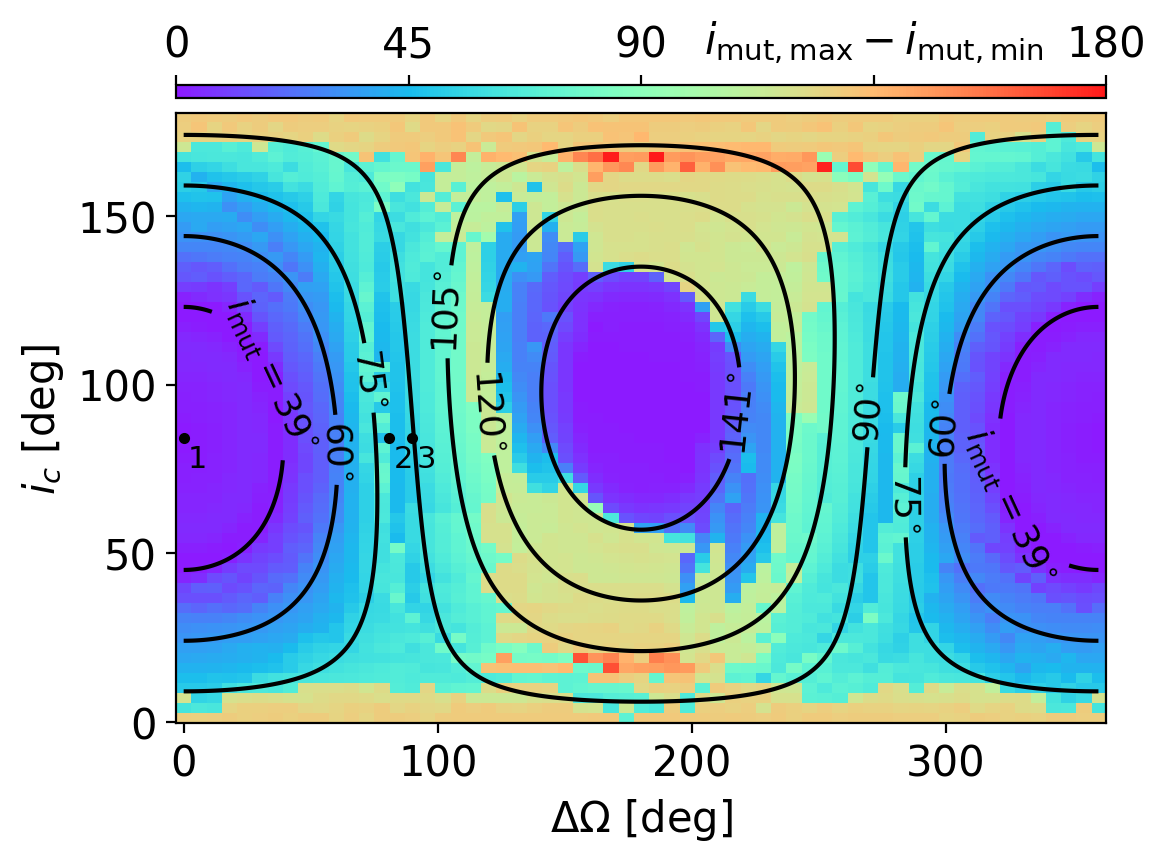}
}
\hbox{
\includegraphics[width=0.333\linewidth]{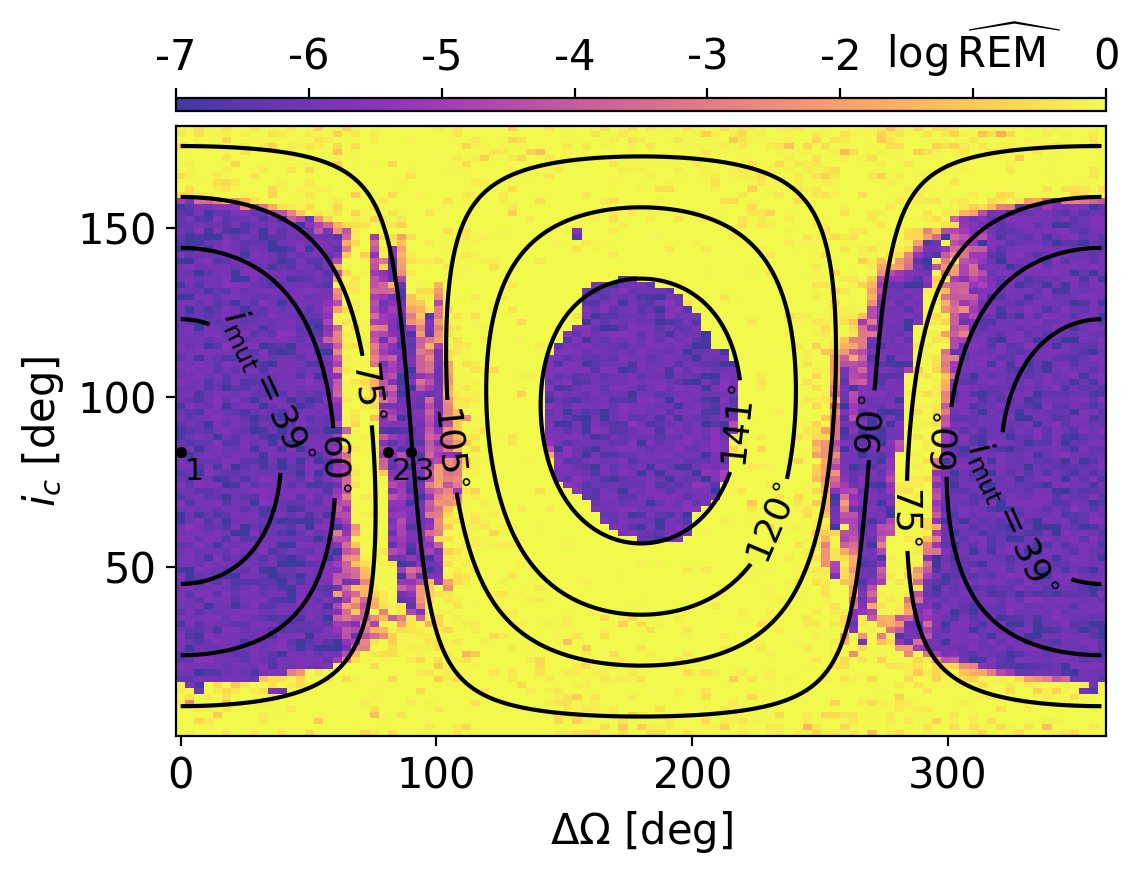}
\includegraphics[width=0.333\linewidth]{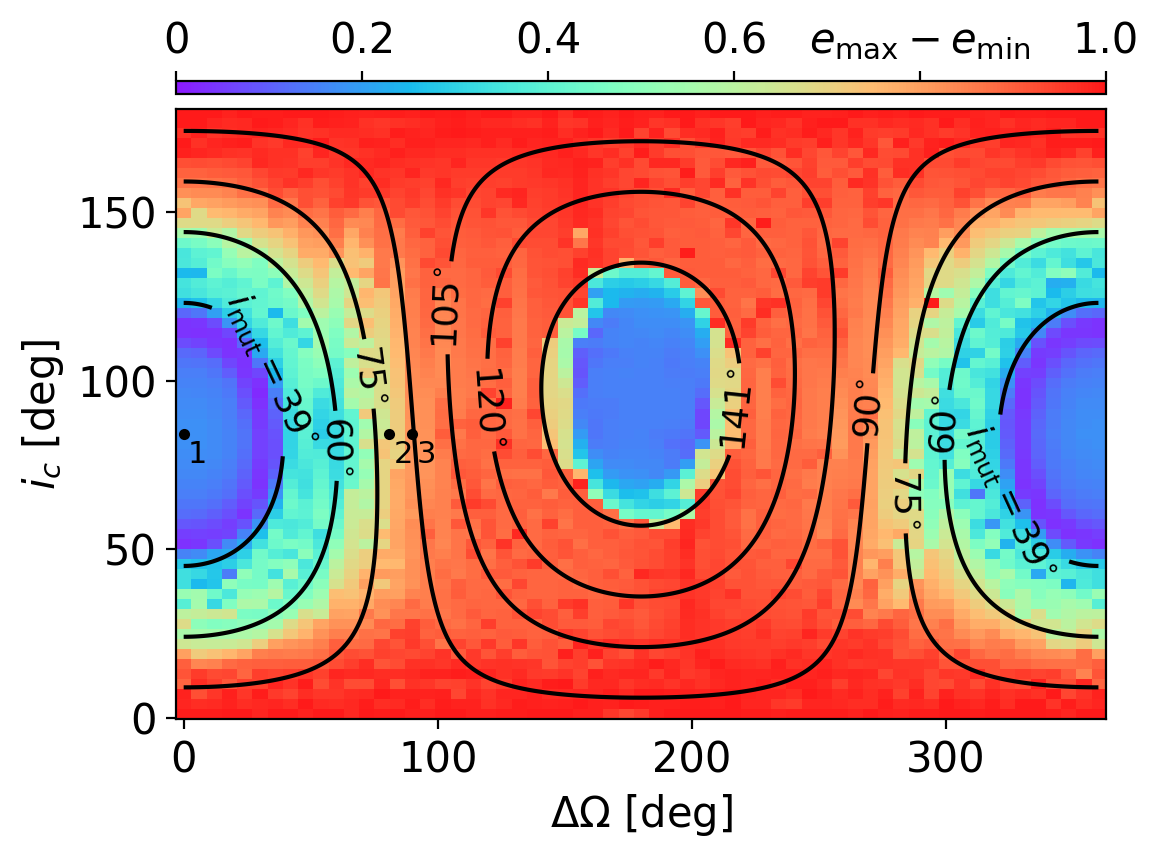}
\includegraphics[width=0.333\linewidth]{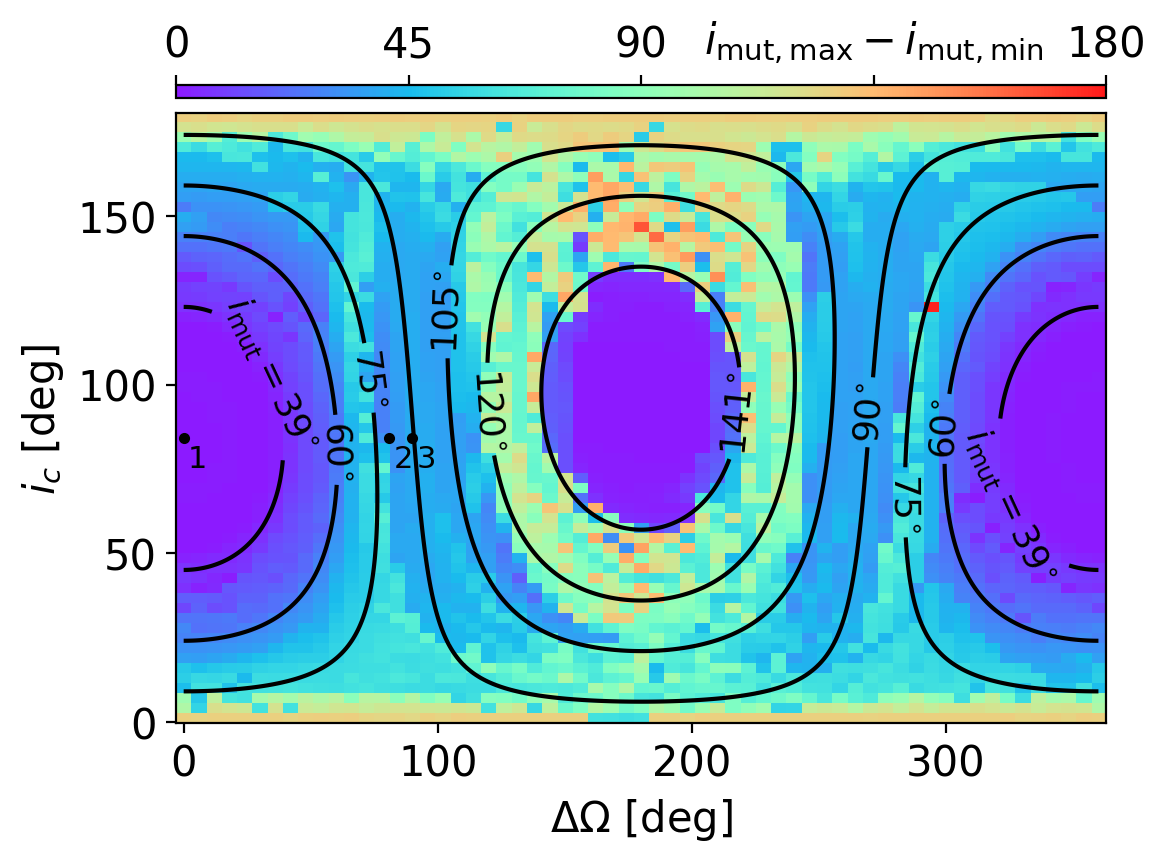}
}
\caption{
Dynamic REM maps (left column), variations of the inner eccentricity (middle column), and mutual inclination of the orbits (right column) for several cycles of oscillation of these elements in the $(\Delta\Omega, i_{c})$ plane. The top row is for the Newtonian, gravitational interactions, the bottom row is for the GR perturbations included in the equations of motion. Black curves mark some levels of mutual inclination of the orbits. The points marked ``1'', ``2'' and ``3'' are for the integrated initial conditions shown in Fig.~\ref{fig:elements}. See text for details.}
\label{fig:mutual}
\end{figure*}

In an attempt to constrain the mutual inclination of the orbits by dynamics, we first computed the REM  maps in the $(\Delta\Omega, i_{c})$--plane, where $\Delta\Omega \in [0^{\circ}, 360^{\circ}]$ is the separation of the nodal lines of the orbits and $i_{c} \in [1^{\circ}, 179^{\circ}]$ is the inclination of the planet~c{}. Then the mutual inclination of the orbits is:
\[
i_{mut} = \cos i_{b} \cos i_{c} + 
\sin i_{b} \cos i_{c} \cos\Delta\Omega. 
\]
The results of the REM simulations are shown in the upper left panel of Fig.~\ref{fig:mutual}. Stable regions appear only in a portion of the parameter plane, outside mutual inclinations approximately $i_{mut} \in (60^{\circ}, 140^{\circ})$. This implies moderate mutual inclinations. Considering that $i_{b} = 84^{\circ} \pm 0.5^{\circ}$, then $i_{c} \simeq (24^{\circ}, 156^{\circ})$ and its mass must be in the range $M_{c} \simeq  (10, 25)$\,M$_{\oplus}$.

\begin{figure*}
\centering
\hbox{
\includegraphics[width=0.333\linewidth]{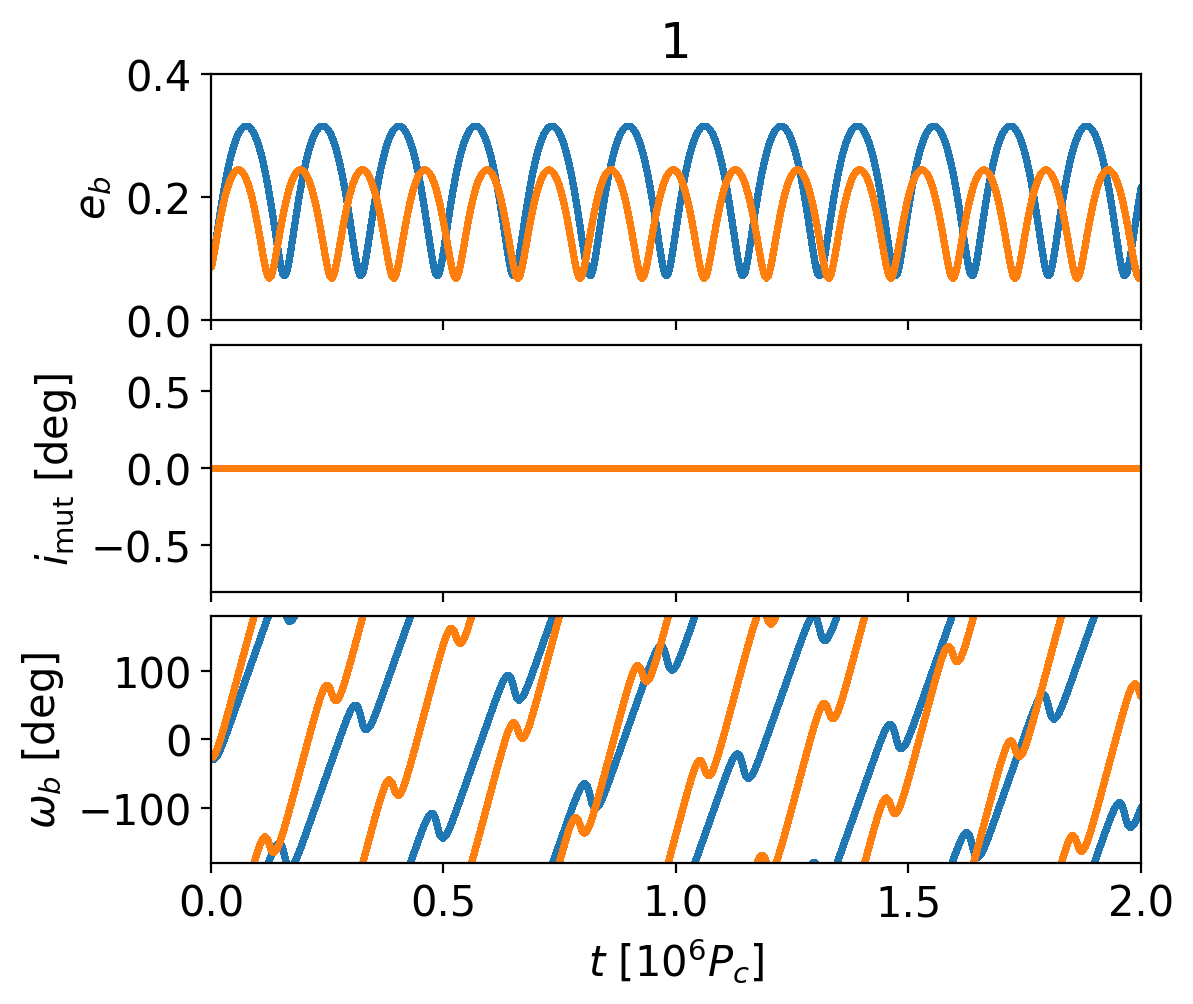}
\includegraphics[width=0.333\linewidth]{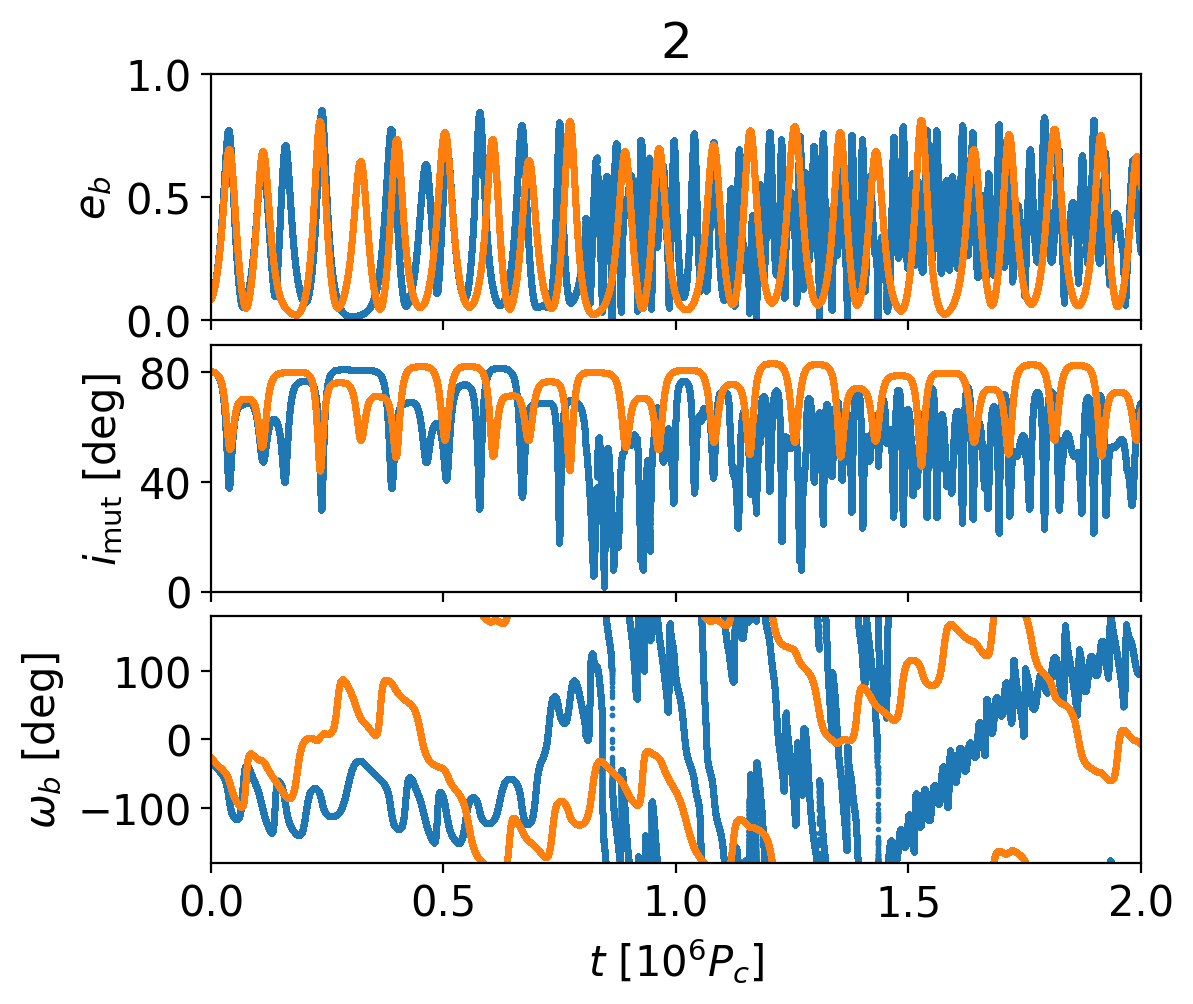}
\includegraphics[width=0.333\linewidth]{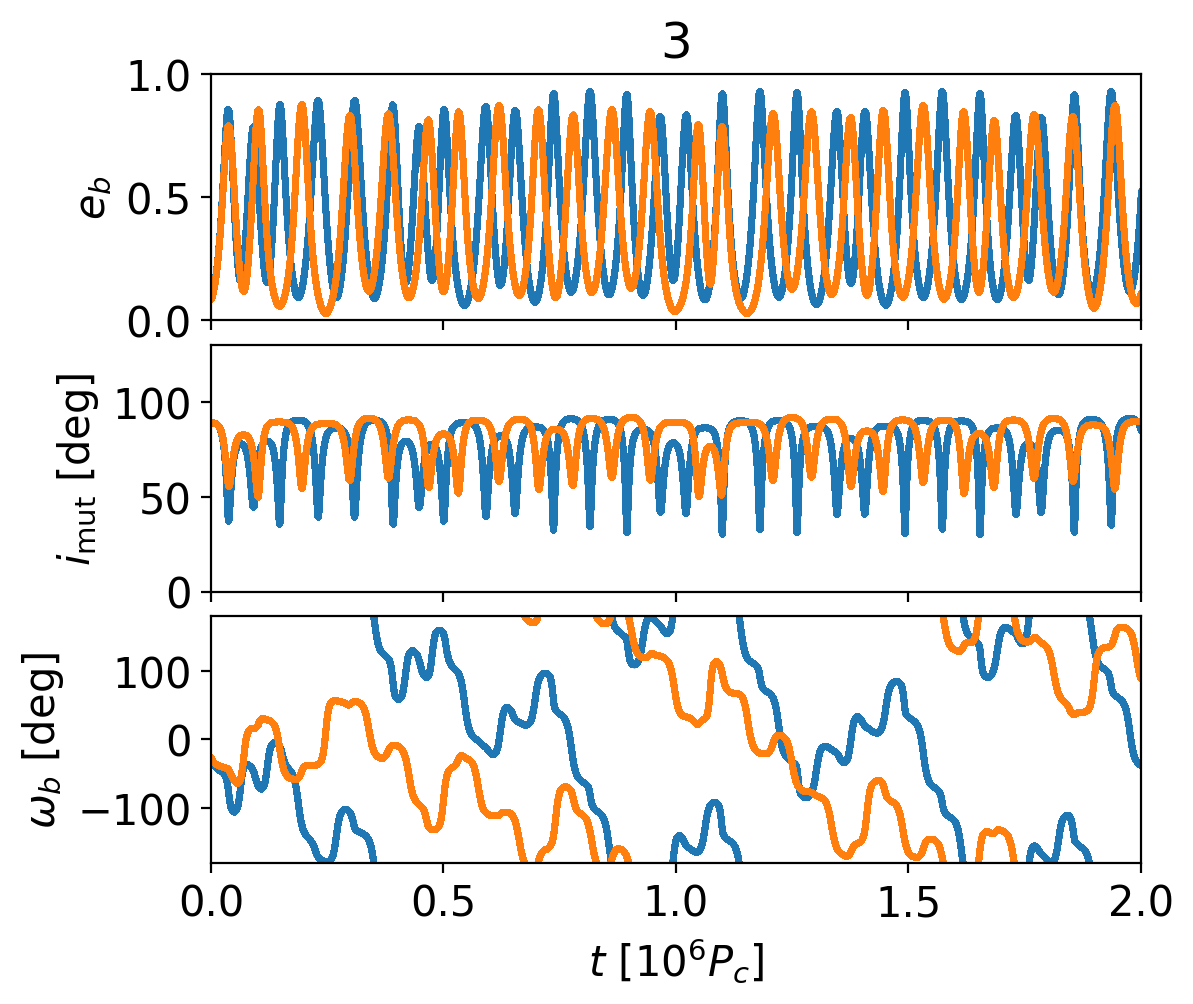}
}
\caption{
The results of the integration of the TOI-2458 system for the best-fitting initial condition with the orbital eccentricity and nodal longitude of the outer orbit marked with ``1'', ``2'' and ``3'' in Fig.~\ref{fig:mutual}. The time evolution of selected orbital elements for the Newtonian force is shown in blue and for the Newtonian and GR models in orange color, respectively. The integrations were performed with the variable-step, high-accuracy \ias{} integrator.}
\label{fig:elements}
\end{figure*}

The resulting scan of the ($\Delta\Omega,i_{c}$)-plane is reminiscent of the results for the Lidov-Kozai resonance \citep[e.g.,][]{Shevchenko2017,Naoz2013}. However, both the nominal system with an initial coplanar orbit and two initial conditions, labeled ``1'', ``2'' and ``3'' respectively, in the REM dynamical map, do not exhibit librations of $\omega_{\rm b}$, so none of them belongs to the libration zone of the resonance. This is illustrated in Fig.~\ref{fig:elements}, where the evolution of the elements has been calculated using the \ias{} integrator of the \rebound{} package, which is also suitable for highly eccentric orbits.  Since the system is relatively compact, with $a_{b}/a_{c} \simeq 0.37$, the presence of a "classical" LK resonance is uncertain. Its analytical models rely on small factor $a_{b}/a_{c}$.  We found numerically that the global dynamics showed features similar to those of the canonical case, such as the large variability of eccentricity and inclination in the anti-phase.

This REM scan is accompanied by panels for the range of eccentricity and mutual inclination of the orbits (the top-middle and top-right panels in Fig.~\ref{fig:mutual} respectively) for the interval spanning at least a few cycles of the variation of the elements shown in Fig.~\ref{fig:elements}. Indeed, in a large part of the phase space, the eccentricity can reach extremely large values, exceeding 0.7-0.8 (the upper-middle panel). This would imply the pericentre distance as small as 0.014--0.0096\,au, only $\simeq 2$ stellar radii of $\simeq 0.006$\,au. At the same time, the mutual inclination in these regions varies up to $\sim 90^{\circ}$ (top right panel). Also, such highly eccentric and mutually inclined systems are usually extremely chaotic and strongly unstable, as can be seen for initial condition ``2'' in Fig.~\ref{fig:elements}, middle panel.

It is also known \citep{Shevchenko2017,Naoz2013} that the LK resonance can be suppressed by General Relativity (GR) perturbations and star-planet tides, which can significantly modify the pericentre precession. We have performed a test of this phenomenon for the GR perturbations in the simplest yet quantitatively correct and CPU-efficient form of the radial and conservative potential $\sim r^{-2}$ introduced by \cite{Nobili1986}. The dynamical map is then substantially modified, with some structures appearing at the boundary of the stable regions (see the lower right panel in Fig.~\ref{fig:mutual}). The eccentricity and mutual inclination regions remain largely similar to the ``pure'' Newtonian case, but some structures, such as the central oval, are significantly different. Clearly, the dynamics of the system are globally affected by the GR perturbation of the Newtonian force.

Moreover, the evolution of the elements for the selected initial conditions shows that the GR perturbation quantitatively modifies the dynamics by changing the frequency of the pericentre precession and the range of the elements in Fig.~\ref{fig:elements}, where orange curves are for the GR model. These simulations also confirm the high sensitivity of REM to chaotic motions and the correct characterization of regular and chaotic solutions in the dynamical maps.

To summarise our results in this section, the overall 3D dynamical structure of the TOI-2458 system is very complex despite apparently well-separated planets with small masses. The results of the dynamical simulations constrain the observed system to relatively small islands of moderate inner, close-in planet eccentricity when the orbital planes are moderately inclined. However, in the case of the hypothetical violent scattering history of the system, we could also observe the orbits just in a phase of increasing or decreasing inner eccentricity. In the case of large variations in this inner eccentricity, the orbital evolution of the inner planet and the whole system would be strongly influenced by the tidal interaction of the star. The properties of the putative LK resonance in the system would then require comprehensive investigation, which is beyond the scope of this paper.

\clearpage

\onecolumn
\section{Additional material}

\begin{longtable}{lccc}
\caption{Median values and 68\% confidence intervals for parameters from the {\tt pyaneti} analysis.} \label{table:planet_par_pyan} \\
\endfirsthead
\multicolumn{3}{l}{Comments: 1 - based on the equation from \citet{Kempton18},}\\
\multicolumn{3}{l}{2 - from stellar parameters, 3 - based on the equation from \citet{Kempton18},}\\
\multicolumn{3}{l}{4 - from $K$ and $R_p/R_\star$, 5 - from planetary parameters,}\\
\multicolumn{3}{l}{6 - based on the equation from \citet{Fossati17}}\\
\endlastfoot
\hline
\hline
Parameter & Unit & Value & Prior\\
\hline
\multicolumn{3}{c}{Input stellar parameters} \\
\hline
$M_\star$ & (M$_{\odot}$) &  $ 1.05 _{- 0.03}^{ + 0.03} $ \\
$R_\star$ & (R$_{\odot}$) & $1.31 _{ - 0.03}^{ + 0.03} $  \\
$\rm T_{eff}$ & ($\mathrm{K}$) & $ 6005 _{- 50}^{ + 50} $ \\
$v_{\rm rot} \sin{i_\star} $ & (km\,s$^{-1}$) & $ 2.5 \pm 2.5 $ \\
$J_{mag}$ & (mag) & $ 8.196 \pm 0.020 $ \\
\hline
\multicolumn{3}{c}{Fitted parameters} \\
\hline
$T_0$ & (\bjdtdb) & $ 2177.4046_{-0.0028}^{+0.0026} $ & $\mathcal{G}$[2177.40,0.01]\\ 
$P$ & (days) &  $ 3.73659_{-0.00047}^{+0.00044} $ & $\mathcal{G}$[3.74,0.01] \\
{$esin\omega$} & () & $ -0.10_{-0.16}^{+0.21} $ & $\mathcal{U}$[-1,1] \\
{$ecos\omega$} & () & $ 0.20_{-0.16}^{+0.11} $ & $\mathcal{U}$[-1,1] \\
b & () & $ 0.847_{-0.042}^{+0.034} $ & $\mathcal{U}$[0,1.2] \\
$\rho_\star$ & (g\,cm$^{-3})$ & $ 0.672_{-0.096}^{+0.93} $ & $\mathcal{G}$[0.66,0.1] \\ 
$R_p/R_\star$ & () & $ 0.0198_{-0.0012}^{+0.0014} $ & $\mathcal{U}$[0,0.1] \\ 
$K$ & (m\,s$^{-1})$ & $ 5.31_{-0.37}^{+0.38} $ & $\mathcal{U}$[0,100] \\
\hline
\multicolumn{3}{c}{Derived parameters for planet TOI-2458\,b} \\
\hline
$M_p$ & (M$_{\oplus}$) & $ 13.31_{-0.96}^{+0.99} $ \\ 
$R_p$ & (R$_{\oplus}$) & $ 2.83_{-0.18}^{+0.20} $ \\
$e$ & () & $ 0.087_{-0.056}^{+0.062} $ \\ 
$\omega$ & (deg) & $ -26.0_{-41.8}^{+57.8} $ \\
$i$ & (deg) & $ 84.03_{-0.52}^{+0.45} $ \\ 
$a$ & (AU) & $ 0.0482_{-0.0027}^{+0.0025} $ \\ 
depth & (ppm) & $ 391.8_{-46.3}^{+55.3} $ \\ 
RM & (m\,s$^{-1}$) & $ 0.31_{-0.11}^{+0.12} $ \\ 
Received irradiance & ($F_{\oplus}$) & $ 865.0_{-77.9}^{+98.8} $ \\ 
Transmission spectroscopy metric$^1$ & () & $ 44.0_{-8.1}^{+10.7} $ \\  
$\rho_\star$$^2$ &(g\,cm$^{-3})$ & $ 0.659_{-0.047}^{+0.051} $ \\ 
$T_{eq}$$^3$ & (K) & $ 1509.4_{-35.2}^{+41.4} $ \\ 
$T_{tot}$ & (hours) & $ 2.13_{-0.14}^{+0.13} $ \\  
$\rho_p$ & (g\,cm$^{-3})$ & $ 3.23_{-0.64}^{+0.76} $ \\  
$g_p$$^4$ & (cm\,s$^{-2}$) & $ 1645_{-293}^{+319} $ \\
$g_p$$^5$ & (cm\,s$^{-2}$) & $ 1632_{-236}^{+264} $ \\
Jeans Escape Parameter$^6$ & () & $ 23.56_{-2.40}^{+2.54} $ \\ 
\hline
\multicolumn{3}{c}{Parameters for planet TOI-2458\,c} \\
\hline
$T_0$ & (\bjdtdb) &  $ 2168.39_{-1.72}^{+1.85} $ & $\mathcal{G}$[2183,7] \\ 
$P$ & (days) &  $ 16.553_{-0.061}^{+0.059} $ & $\mathcal{G}$[16.6,1.4] \\
{$esin\omega$} & () & $ 0.04_{-0.32}^{+0.28} $ & $\mathcal{U}$[-1,1] \\
{$ecos\omega$} & () & $ 0.45_{-0.36}^{+0.17} $ & $\mathcal{U}$[-1,1] \\
$K$ & (m\,s$^{-1})$ & $ 2.63_{-0.48}^{+0.48} $ & $\mathcal{U}$[0,100] \\
\hline
\multicolumn{3}{c}{Derived parameters for planet TOI-2458\,c} \\
\hline
$M_psini$ & (M$_{\oplus}$) & $ 10.22_{-1.90}^{+1.71} $\\
e & () & $ 0.29 \pm 0.16 $ \\ 
$\omega$ & (deg) & $ 4.3_{-41.1}^{+48.1} $ \\ 
$a$ & (AU) & $ 0.129_{-0.002}^{+0.002} $ \\
\hline
\multicolumn{3}{c}{Other parameters} \\
\hline
$q_1$ & () & $ 0.43_{-0.28}^{+0.36} $ & $\mathcal{U}$[0,1]\\ 
$q_2$ & () & $ 0.37_{-0.25}^{+0.30} $ & $\mathcal{U}$[0,1]\\ 
$u_1$ & () & $ 0.44_{-0.31}^{+0.36} $ & $\mathcal{U}$[0,1]\\ 
$u_2$ & () & $ 0.14_{-0.31}^{+0.39} $ & $\mathcal{U}$[0,1]\\ 
Sys. vel. HARPS RVs & (km\,s$^{-1}$) & $ 0.0016_{-0.0041}^{+0.0049} $ & $\mathcal{U}$[-0.5,0.5] \\ 
Sys. shift HARPS S-index & () & $ 0.1544 \pm 0.0012 $ & $\mathcal{U}$[-0.5,0.5] \\
Sys. shift HARPS H$\alpha$ & () & $ 2.15_{-0.20}^{+0.24} $ & $\mathcal{U}$[-0.5,0.5] \\
RV jitter & (m\,s$^{-1}$) & $ 1.67_{-0.29}^{+0.35} $ & $\mathcal{U}$[0,5]\\
Transit jitter & () & $ 0.0005144 \pm 0.0000032 $ & $\mathcal{U}$[0,0.005]\\
\smallskip\\
\hline
\multicolumn{3}{c}{GP parameters} \\
\hline
A0 (RV) & (km\,s$^{-1}$) & $ 0.0084_{-0.0040}^{+0.0053} $ & $\mathcal{U}$[-0.5,0.5] \\
A1 (RV) & (km\,s$^{-1}$d) & $ -0.051_{-0.045}^{+0.031} $ & $\mathcal{U}$[-0.5,0.5] \\
A2 (S-index) &  & $ -0.0007_{-0.0030}^{+0.0024} $ & $\mathcal{U}$[-0.5,0.5] \\
A3 (S-index) &  & $ -0.123_{-0.065}^{+0.053} $ & $\mathcal{U}$[-0.5,0.5] \\
$\lambda_e$ & (days) & $ 74.7_{-25.7}^{+17.6} $ & $\mathcal{U}$[1,100] \\
$\lambda_p$ & & $ 3.10_{-1.18}^{+1.24} $ & $\mathcal{U}$[0.1,5] \\
P$_{GP}$ & (days) & $ 47.97_{-4.02}^{+6.96} $ & $\mathcal{U}$[40,60] \\
\hline
\hline

\smallskip\\
\end{longtable}

\onecolumn

\begin{longtable}{lcccc}
\caption{Relative HARPS radial velocities and activity indicators of TOI-2458.} \label{tab:long} \\

\hline \multicolumn{1}{c}{\textbf{Date (BJD)}} & \multicolumn{1}{c}{\textbf{RV (m/s)}} & \multicolumn{1}{c}{\textbf{$\sigma_{RV}$ (m/s)}} & 
\multicolumn{1}{c}{\textbf{H$\alpha$}} & 
\multicolumn{1}{c}{\textbf{S-index}} \\ \hline 
\endfirsthead

\multicolumn{5}{c}%
{{\bfseries \tablename\ \thetable{} -- continued from previous page}} \\
\hline \multicolumn{1}{c}{\textbf{Date (BJD)}} & \multicolumn{1}{c}{\textbf{RV (m/s)}} & \multicolumn{1}{c}{\textbf{$\sigma_{RV}$ (m/s)}} & \multicolumn{1}{c}{\textbf{H$\alpha$}} & \multicolumn{1}{c}{\textbf{S-index}} \\ \hline 
\endhead

\hline \multicolumn{4}{r}{{Continued on next page}} \\ 
\endfoot

\hline 
\endlastfoot

2459626.57046 & -3.60580 & 1.74066 & 1.06960 & 0.15825\\
2459627.57121 & 0.45331 & 1.63245 & 1.07206 & 0.15971\\
2459628.52387 & 9.23207 & 1.30024 & 1.07443 & 0.15460\\
2459628.59374 & 8.08026 & 1.46562 & 1.07478 & 0.16047\\
2459629.52190 & 5.98527 & 1.24333 & 1.07262 & 0.15703\\
2459629.59192 & 3.80151 & 1.46576 & 1.07208 & 0.16234\\
2459630.52217 & -4.23015 & 1.24629 & 1.06989 & 0.16065\\
2459630.56781 & -4.25610 & 1.14161 & 1.07355 & 0.15881\\
2459630.58519 & -4.79558 & 1.15478 & 1.07014 & 0.15796\\
2459631.52037 & 0.00000 & 1.33286 & 1.06902 & 0.15526\\
2459631.58935 & 2.23137 & 1.30985 & 1.07101 & 0.15920\\
2459632.52137 & 6.28987 & 1.26168 & 1.07227 & 0.15689\\
2459632.60050 & 3.26533 & 1.33886 & 1.06858 & 0.16014\\
2459633.52765 & -5.03571 & 1.34001 & 1.07016 & 0.15679\\
2459633.60602 & -7.18474 & 1.51795 & 1.06998 & 0.15860\\
2459634.52241 & -7.48829 & 1.57321 & 1.06951 & 0.15818\\
2459634.60253 & -9.21190 & 1.61501 & 1.06998 & 0.15735\\
2459636.51653 & 0.96106 & 1.23570 & 1.07155 & 0.15686\\
2459636.59387 & -0.08565 & 1.32675 & 1.06968 & 0.15648\\
2459637.51561 & -4.62772 & 1.40295 & 1.07042 & 0.15445\\
2459637.59235 & -7.00802 & 1.69053 & 1.07368 & 0.16017\\
2459638.51472 & -6.17653 & 1.22186 & 1.07361 & 0.15664\\
2459639.54902 & 1.42886 & 1.39532 & 1.07361 & 0.15683\\
2459640.51211 & -1.84278 & 1.70393 & 1.07007 & 0.16079\\
2459641.51016 & -8.19887 & 1.30513 & 1.07639 & 0.15843\\
2459642.50970 & -9.35474 & 1.54382 & 1.07154 & 0.15811\\
2459643.51141 & -0.79903 & 1.46698 & 1.07540 & 0.15463\\
2459644.50853 & -3.35652 & 2.04471 & 1.07520 & 0.14964\\
2459645.50755 & -7.06186 & 1.75046 & 1.07355 & 0.14731\\
2459646.50805 & -3.38800 & 1.55573 & 1.07195 & 0.15289\\
2459647.50417 & 0.40703 & 1.50461 & 1.07209 & 0.15054\\
2459648.50484 & -6.72600 & 1.27400 & 1.07290 & 0.15213\\
2459653.53430 & -5.78354 & 1.54562 & 1.07807 & 0.14714\\
2459657.50361 & -5.01645 & 1.39315 & 1.07207 & 0.14690\\
2459657.53415 & -6.14540 & 1.53096 & 1.07699 & 0.14837\\
2459658.49814 & 2.95163 & 1.45863 & 1.07256 & 0.14987\\
2459659.49560 & -0.19182 & 1.56236 & 1.07232 & 0.14801\\
2459660.49647 & -3.01936 & 1.57569 & 1.07350 & 0.15198\\
2459661.49501 & 8.11929 & 1.52942 & 1.07258 & 0.14718\\
2459662.49028 & 13.95170 & 1.69995 & 1.07152 & 0.14842\\
2459664.52268 & -3.64051 & 1.36848 & 1.07040 & 0.14676\\
2459665.48892 & 6.66790 & 1.36021 & 1.07136 & 0.15873\\
2459666.52975 & 3.12895 & 1.60524 & 1.07196 & 0.15846\\
2459668.48463 & -3.76043 & 1.61613 & 1.06904 & 0.15566\\
2459670.48430 & -0.90826 & 1.56124 & 1.06785 & 0.16045\\
2459671.49321 & -1.32617 & 1.35500 & 1.07331 & 0.15665\\
2459672.48851 & -2.25176 & 1.60849 & 1.07266 & 0.15310\\
2459674.47901 & 0.67112 & 1.48633 & 1.07033 & 0.15450\\
2459675.47852 & 2.34194 & 1.56661 & 1.06851 & 0.15729\\
2459677.48053 & 10.91969 & 1.50582 & 1.07440 & 0.15440\\
2459678.47977 & -0.61313 & 1.69065 & 1.07390 & 0.15629\\
2459680.47754 & 10.76232 & 2.11955 & 1.07010 & 0.15827\\
2459681.47464 & 8.98306 & 1.82464 & 1.06534 & 0.15895\\
2459682.47761 & -5.07983 & 1.91448 & 1.06723 & 0.15419\\
2459685.48173 & 0.63113 & 1.60358 & 1.06884 & 0.15533\\
2459891.76265 & 6.06616 & 1.21894 & 1.07393 & 0.15666\\
2459893.73163 & 14.80380 & 0.84291 & 1.07371 & 0.15849\\
2459898.70094 & 5.79629 & 1.20872 & 1.07310 & 0.15829\\
2459900.74072 & 3.32741 & 1.69840 & 1.06977 & 0.15519\\
2459901.79418 & 5.58911 & 1.06385 & 1.07311 & 0.15309\\
2459902.72177 & -0.96515 & 1.50172 & 1.07380 & 0.15649\\
2459904.71755 & 6.31906 & 1.61721 & 1.07318 & 0.15806\\
2459906.74864 & -0.56126 & 1.62528 & 1.07719 & 0.15355\\
2459907.70776 & 6.29759 & 1.79251 & 1.07733 & 0.15792\\
2459908.72213 & 10.87003 & 2.18669 & 1.07902 & 0.14902\\
2459909.75548 & 6.43331 & 1.29718 & 1.07080 & 0.15772\\
2459927.66966 & 12.34636 & 2.11303 & 1.07640 & 0.15161\\
2459929.67684 & -5.51379 & 1.79868 & 1.06982 & 0.15410\\
2459931.72420 & -2.06245 & 1.78946 & 1.06959 & 0.15507\\
2459936.59670 & -6.77609 & 1.69414 & 1.07239 & 0.15140\\
2459940.56949 & -6.34509 & 1.57602 & 1.07320 & 0.15196\\
2459948.72207 & 0.43391 & 1.70429 & 1.07507 & 0.15224\\
2459972.56829 & 6.36334 & 1.39499 & 1.07297 & 0.15641\\
2459975.60344 & 2.94834 & 1.47080 & 1.07454 & 0.15427\\
2459978.64380 & -5.00263 & 2.13103 & 1.07260 & 0.15359\\
2459981.59711 & -1.95722 & 2.25326 & 1.07387 & 0.15369\\
2459983.64077 & 3.23188 & 2.27755 & 1.07863 & 0.14415\\
2460017.54799 & 0.59316 & 2.02706 & 1.07640 & 0.15167\\
2460019.54102 & -5.88262 & 1.82536 & 1.07755 & 0.15345\\
2460021.53921 & 0.67872 & 1.53177 & 1.07248 & 0.15617\\
2460025.52714 & 3.98090 & 1.57371 & 1.07558 & 0.15725\\
2460029.49171 & 3.00734 & 1.90912 & 1.07559 & 0.14690\\
2460033.49697 & -3.12350 & 1.60431 & 1.07157 & 0.15176\\
2460035.48864 & 2.72208 & 1.81145 & 1.06996 & 0.15296\\
2460178.88692 & -0.42893 & 2.66881 & 1.07621 & 0.15933\\
2460215.85413 & 1.04048 & 2.23979 & 1.07201 & 0.15809\\
\hline

\end{longtable}

\begin{figure*}[ht!]
\centering
\includegraphics[width=0.85\textwidth]{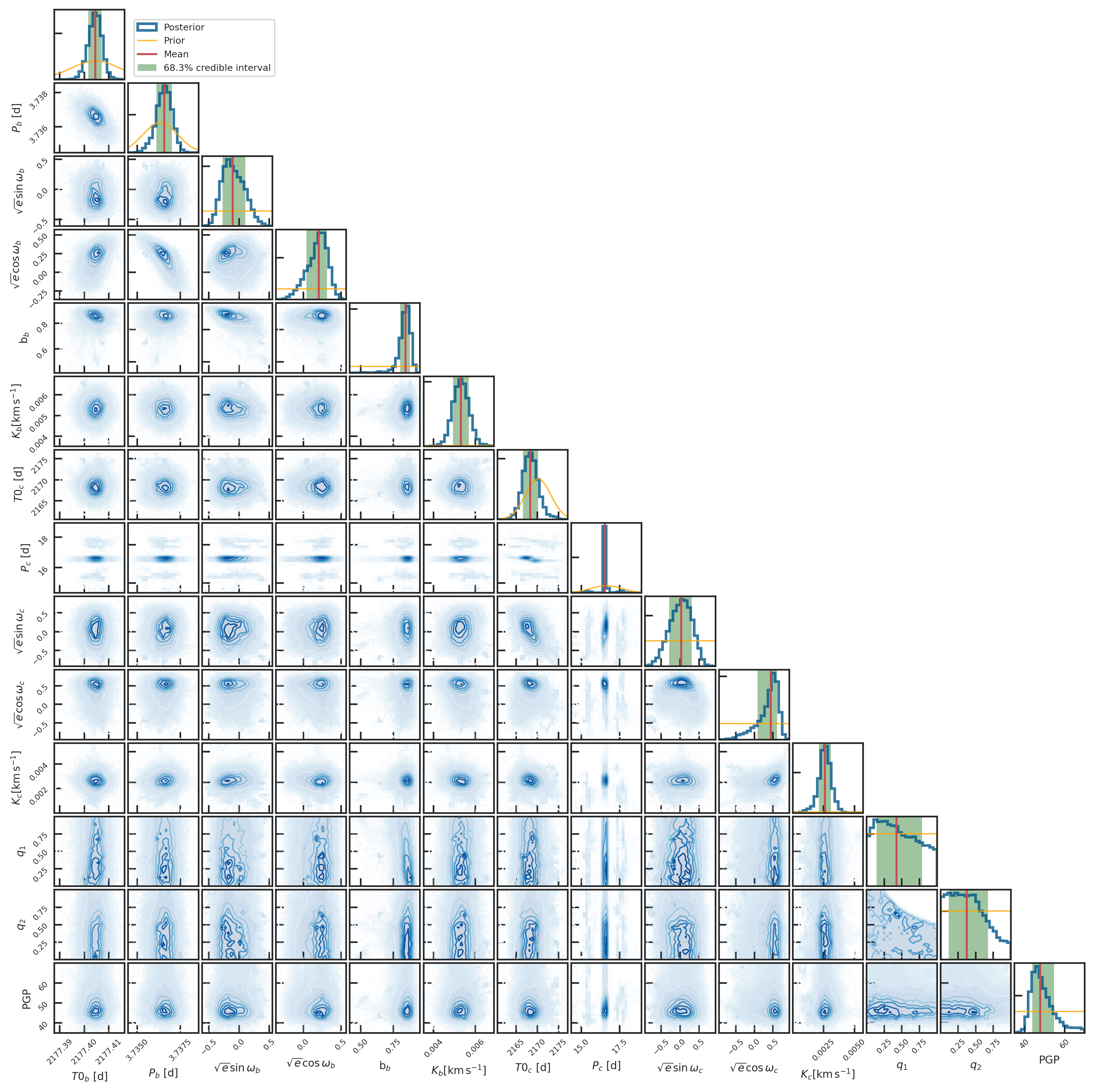}
\caption{Correlations between the main free parameters of the {\tt pyaneti} model from the MCMC sampling. At the end of each row is shown the derived posterior probability distribution.} \label{fig:corr}
\end{figure*}

\end{document}